\documentclass[longauth]{aa} 

\usepackage{graphicx}
\usepackage{txfonts}
\usepackage{natbib}

\def\e20{$\times 10^{20}$}

\def\10K{10K-sample} 
\def\20K{20K-sample} 
\def\kmsec{km s$^{-1}$}
\def\kmsecmpc{km s$^{-1} Mpc^{-1}$}
\def\Mpch{$h^{-1}_{70}~Mpc$~}
\def\ie{i.e.,~}
\def\vs{versus~}
\def\eg{e.g.,~}

\def\bluefrac{$\it{F_{blue}}$~}
\def\today{\number\day -\number\month -\number\year}

\begin{document}

\title{The zCOSMOS Redshift Survey: \\                            
How group environment alters global downsizing trends
\thanks{based on data obtained with the European Southern Observatory 
Very Large Telescope, Paranal, Chile, program 175.A-0839, PI: S. Lilly}}

\author{
A.~Iovino\inst{1}
\and
O.~Cucciati\inst{1,2}
\and
M.~Scodeggio\inst{3}
\and 
C.~Knobel\inst{4}
\and
K.~Kova\v{c}\inst{4}
\and
S.~Lilly\inst{4}
\and
M.~Bolzonella\inst{5}
\and 
L.~A.~M.~Tasca\inst{2,3}
\and
G.~Zamorani\inst{5}
\and
E.~Zucca\inst{5}
\and 
K.~Caputi\inst{4}
\and
L.~Pozzetti\inst{5}
\and 
P.~Oesch\inst{4}
\and
F.~Lamareille\inst{6}
\and
C.~Halliday\inst{7}
\and
S.~Bardelli\inst{5}
\and 
A.~Finoguenov\inst{8}
\and 
L.~Guzzo\inst{1}
\and
P.~Kampczyk\inst{4}
\and
C.~Maier\inst{4}
\and
M.~Tanaka\inst{9}
\and
D.~Vergani\inst{5}
\and  
C.~M.~Carollo\inst{4}
\and
T.~Contini\inst{6}
\and
J.-P.~Kneib\inst{2}
\and
O.~Le~F\`{e}vre\inst{2}
\and
V.~Mainieri\inst{9}
\and
A.~Renzini\inst{10}
\and  
A.~Bongiorno\inst{8}
\and
G.~Coppa\inst{5}
\and
S.~de~la~Torre\inst{2,1}
\and
L.~de~Ravel\inst{2}
\and
P.~Franzetti\inst{3}
\and
B.~Garilli\inst{3}
\and
J.-F.~Le~Borgne\inst{6}
\and
V.~Le~Brun\inst{2}
\and
M.~Mignoli\inst{5}
\and 
R.~Pell\`o\inst{6}
\and
Y.~Peng\inst{4}
\and
E.~Perez-Montero\inst{6}
\and
E.~Ricciardelli\inst{10}
\and
J.~D.~Silverman\inst{4}
\and
L.~Tresse\inst{2}
\and  
U.~Abbas\inst{11}
\and
D.~Bottini\inst{3}
\and
A.~Cappi\inst{5}
\and
P.~Cassata\inst{2,12}
\and
A.~Cimatti\inst{13}
\and 
A.~M.~Koekemoer\inst{15}
\and 
A.~Leauthaud \inst{14}
\and 
D.~Maccagni\inst{3}
\and 
C.~Marinoni\inst{16}
\and 
H.~J.~McCracken\inst{17}
\and 
P.~Memeo\inst{3}
\and 
B.~Meneux\inst{3,18}
\and 
C.~Porciani\inst{4,19}
\and 
R.~Scaramella\inst{20}
\and 
D.~Schiminovich\inst{21}
\and  
N.~Scoville\inst{22}
}

\offprints{Angela Iovino (angela.iovino@brera.inaf.it)}

\institute{
INAF - Osservatorio Astronomico di Brera, Via Brera, 28, I-20159 Milano, Italy \\ 
\email{angela.iovino@brera.inaf.it}
\and
{Laboratoire d'Astrophysique de Marseille, CNRS-Universit{\'e} d'Aix-Marseille, 38 rue Frederic Joliot Curie, F-13388 Marseille, France}  
\and
{INAF - IASF Milano, Via Bassini 15, I-20133, Milano, Italy}  
\and
{Institute of Astronomy, ETH Zurich, CH-8093, Z\"urich, Switzerland}  
\and
{INAF - Osservatorio Astronomico di Bologna, via Ranzani 1, I-40127 Bologna, Italy}  
\and
{Laboratoire d'Astrophysique de Toulouse-Tarbes, Universit\'{e} de Toulouse, CNRS, 14 avenue Edouard Belin, F-31400 Toulouse, France}  
\and 
{INAF - Osservatorio Astrofisico di Arcetri, Largo Enrico Fermi 5, I - 50125 Firenze, Italy}  
\and 
{Max-Planck-Institut f\"ur extraterrestrische Physik, D-84571 Garching b. Muenchen, D-85748, Germany}  
\and
{European Southern Observatory, Karl-Schwarzschild-Strasse 2, Garching b. Muenchen, D-85748, Germany}  
\and
{Dipartimento di Astronomia, Universit\`a di Padova, vicolo Osservatorio 3, I-35122 Padova, Italy}  
\and
{INAF - Osservatorio Astronomico di Torino, Strada Osservatorio 20, I-10025 Pino Torinese, Torino, Italy}  
\and 
{Dept. of Astronomy, University of Massachusetts, 710 North Pleasant Street, Amherst, MA 01003-9305, USA}  
\and 
{Dipartimento di Astronomia, Universit\'a di Bologna, via Ranzani 1, I-40127, Bologna, Italy}  
\and
{Physics Division, MS 50 R5004, Lawrence Berkeley National Laboratory, 1 Cyclotron Rd., Berkeley, CA 94720, USA}  
\and 
{Space Telescope Science Institute, 3700 San Martin Drive, Baltimore, MD 21218, USA}   
\and 
{Centre de Physique Theorique, Campus de Luminy, Case 907 - F-13288 Marseille, France}  
\and
{Instit}ut d'Astrophysique de Paris, UMR 7095 CNRS, Universit\'e Pierre et Marie Curie, 98bis boulevard Arago, F-75014 Paris, France  
\and 
Universit\"ats-Sternwarte, Scheinerstrasse 1, Munich D-81679, Germany 
\and 
{Argelander-Institut f\"{u}r Astronomie, Auf dem H\"{u}gel 71, D-53121 Bonn, Germany}  
\and
{INAF, Osservatorio di Roma Via di Frascati, 33, I-00040 Monte Porzio Catone, Italy}  
\and
{Department of Astronomy, Columbia University, 550 West 120th Street, New York, NY 10027, USA}  
\and
{California Institute of Technology,  MC 105-24, 1200 East California Boulevard, Pasadena, CA 91125, USA}  
}
\date{Received date - ; accepted date - }

\abstract 
{Groups of galaxies are a common environment, bridging the gap between
starforming field galaxies and quiescent cluster galaxies.  Within
groups secular processes could be at play, contributing to the
observed strong decrease of star formation with cosmic time in the
global galaxy population.}
{We took advantage of the wealth of information provided by the first
$\sim 10000$ galaxies of the zCOSMOS-bright survey and its group
catalogue to study in detail the complex interplay between group
environment and galaxy properties. }
{The classical indicator $\it{F_{blue}}$, \ie the fraction of blue galaxies,
proved to be a simple but powerful diagnostic tool.  We studied its
variation for different luminosity and mass selected galaxy samples,
divided as to define groups/field/isolated galaxy subsamples.}
{Using rest-frame evolving B-band volume-limited samples, the groups
galaxy population exhibits significant blueing as redshift increases, but
maintains a systematic difference (a lower $\it{F_{blue}}$) with respect to
the global galaxy population, and an even larger difference with
respect to the isolated galaxy population.  However moving to mass
selected samples it becomes apparent that such differences are largely
due to the biased view imposed by the B-band luminosity selection,
being driven by the population of lower mass, bright blue galaxies for
which we miss the redder, equally low mass, counterparts. By carefully
focusing the analysis on narrow mass bins such that mass segregation
becomes negligible we find that only for the lowest mass bin explored,
\ie $log({\cal M}_{*}/{\cal M}_{\odot}) \leq 10.6 $, does a significant
residual difference in color remain as a function of environment,
while this difference becomes negligible toward higher masses.}
{Our results indicate that red galaxies of mass $log({\cal
M}_{*}/{\cal M}_{\odot}) \geq 10.8 $ are already in place at $z \sim
1$ and do not exhibit any strong environmental dependence, possibly
originating from so-called nature or internal mechanisms. In contrast,
for lower galaxy masses and redshifts lower than $z\sim 1$, we observe
the emergence in groups of a population of nurture red galaxies:
slightly deviating from the trend of the downsizing scenario followed
by the global galaxy population, and more so with cosmic time. These
galaxies exhibit signatures of group-related secular physical
mechanisms directly influencing galaxy evolution. Our analysis implies
that these mechanisms begin to significantly influence galaxy
evolution after $z \sim 1$, a redshift corresponding to the emergence of
structures in which these mechanisms take place.}


\authorrunning {A.~Iovino et al.}
\titlerunning {The zCOSMOS Redshift Survey: how group environment alters global downsizing trends}

\maketitle

\section{Introduction}
\label{sect:Intro} 

Groups and clusters are commonly viewed as sites where environmental
influences can affect the colors, star formation histories and
morphologies of their member galaxies.  One of the first pieces of
empirical evidence supporting this claim was the observation by
Butcher \& Oemler that clusters of galaxies contain a higher fraction
of blue galaxies at progressively higher redshift, the so-called
Butcher-Oemler effect
\citep{ButcherOemler1978,ButcherOemler1984}. Their result provided
direct observational evidence of strong, rapidly evolving galaxy
population colors inside cluster cores with redshift.

Since these early papers, the Butcher-Oemler effect has been confirmed
photometrically \citep{Rakos1995, Margoniner2000, Margoniner2001,
Kodama2001, Goto2003}, spectroscopically \citep{DresslerGunn1982,
DresslerGunn1992, Lavery1986, Lavery1988, Fabricant1991,
Poggianti1999, Poggianti2006, Ellingson2001}, has been extended to
groups \citep{Allington-Smith1993, Wilman2005b, Gerke2007,
Cucciati2009a}, and critically discussed in the context of selection
biases \citep{Andreon1999, Andreon2004, Andreon2006}.

In parallel to these studies, evidence has emerged that the Universe
as a whole formed stars more actively in the past than today
\citep{Lilly1996, Madau1998, Hopkins2004, Schiminovich2005} and that
the typical mass of galaxies where the bulk of star formation occurs
is higher in the past than today, the so-called downsizing effect
\citep{Cowie1996, Gavazzi1996}.

These observations questioned whether the Butcher-Oemler phenomenon
is caused by physical mechanisms typical of dense environments, that
significantly alter global trends displayed by the global galaxy
population in the coeval field, or reflects the evolution of
the global galaxy population. Interestingly, with growing evidence
that denser environments only suppress star formation
\citep{Balogh2004a, Balogh2004b}, we have started to test in groups
and/or clusters at higher redshifts whether we measure a higher
fraction of blue galaxies that is nevertheless lower than the coeval
field.

Since groups contain a large fraction of galaxies in the nearby
Universe, nearly $\sim 50\%$ \citep{HuchraGeller1982, Eke2004,
Berlind2006}, while only a few percent of galaxies are contained in
the denser cluster cores, group-related transformations may drive the
observed strong decrease in star formation with cosmic time, at least
since $z \sim 1$, when these structures started to become predominant
according to the hierarchical structure scenarios.

In the cores of rich clusters phenomena such as ram pressure stripping
have been widely documented in the literature, as for well studied
galaxies in the Virgo cluster, \citep{Kenney2004, Vollmer2004}, and
observed in simulations \citep{Bruggen2008}.

In contrast, similar environment-dependent effects in groups are less
clearly defined, although possibilities have been presented in the
literature, including gradual cessation of star formation induced either by
gentle gas stripping and starvation by a diffuse intragroup medium, or
by slow group-scale harassment \citep{Larson1980, Moore1999,
Gnedin2003, Roediger2005, Kawata2008}.

From a theoretical perspective, numerical simulations incorporating the
standard cosmological paradigm suggest that galaxy properties such as,
\eg colors, spin etc.) are primarily determined by the mass of the
dark matter halo in which the galaxy resides \citep{CooraySheth2002},
and that, at a given mass, in overdense environments dark matter
haloes assemble at higher redshifts than in underdense environments
\citep{Gao2005}. This framework could provide a simple way of
explaining the observed trends in colors with galaxy luminosity, mass
and environment, at low \citep{DePropris2004, BlantonBerlind2007} and
also at high redshifts \citep{Wilman2005b, Balogh2007}, without
resorting to any specific mechanisms acting in groups.  Two large
recent redshift surveys, VVDS and DEEP2, have addressed this problem
by studying both groups \citep{Gerke2007, Cucciati2009a} and local
density field measurements \citep{Cucciati2006, Cooper2006,
Cooper2007}, although both studies considered both luminosity selected
samples, a choice that, as we discuss later, offers only partial
insight into the problem. 

The question of which variables are needed to fully define galaxy
evolution therefore remains unanswered, and is usually considered
in terms of either {\it nature} or {\it nurture} processes.  This
corresponds to asking whether: galaxy evolution is driven mainly by
internal processes, imprinted at galaxy birth, that operate inside the
average galaxy, or group environment has a specific effect on shaping
galaxy evolution, because of specific mechanisms taking place in
dense, possibly virialized regions, where secular influences have
better chances to affect galaxy evolution.

To distinguish between the effects of environment and trends related
to galaxy evolution with redshift, one needs homogeneous and sizeable
group and field galaxy samples, covering a wide redshift range and
with reliable measurements of galaxy rest-frame colors, luminosities
and masses. These data would allow us to monitor with look-back time
the evolutionary histories of galaxies located in different
group/field environments, and to disentangle between the different
dependencies and their relative importance.

The advantage of the data-set used in our analysis is that it
satisfies all of these requirements. zCOSMOS is a survey tailored for
studying the large scale structure and detecting groups up to $z \sim
1$ \citep{Lilly2007, Lilly2009}. Its large volume coverage and small
errors in galaxy redshift measurements enable the production, even for
the first batch of $\sim 10000$ measured redshifts, of a large group
catalogue containing $102$ groups with $N \geq 5$ spectroscopically
confirmed members and a further $\sim 700$ going down to pairs
\citep{Knobel2009}. Furthermore this catalogue, because of the precise
fine tuning of the algorithm used for group detection, is remarkably
free from contamination and incompleteness, especially at the low
richness, low velocity dispersion end and, most importantly, its
quality is stable as a function of redshift \citep{Knobel2009}.  Last
but not least, the large amount of precise photometric ancillary data
available from the COSMOS survey \citep{Scoville2007}, provides robust
estimates of the fundamental properties of each galaxy, such as
rest-frame luminosities, colors and masses. We are therefore in best
position with our data-set to investigate in detail which processes
are the most influential in shaping galaxy evolution.

Complementary analyses of the same \10K data-set have been carried out
in other papers.  \citet{Kovac2009b} in a parallel paper study the
influence of group environment in shaping galaxy morphologies.  Using
the density field measured in \citet{Kovac2009a}, \citet{Zucca2009}
and \citet{Bolzonella2009} study the galaxy luminosity and mass
functions respectively as a function of environment, while
\citet{Cucciati2009b} and \citet{Tasca2009} investigate the
dependencies of galaxy colors and morphologies, respectively, from the
general density field.  Finally \citet{Silverman2009} and
\citet{Vergani2009} study how environment plays a role in triggering
active galactic nuclei activity and in quenching star-formation
respectively. For more details, we refer the interested reader to
those papers.

A concordance cosmology is adopted throughout our paper, with $h_{70}
= H_0/70$ \kmsecmpc, $\Omega_{m} = 0.25$ and $\Omega_{\Lambda } =
0.75$.  All magnitudes are always quoted in the AB system.

\section{Samples used in the analysis}

\subsection{The zCOSMOS \10K} 
\label{sect:sample10K} 
 
The zCOSMOS survey is a large spectroscopic survey undertaken in the
COSMOS field \citep{Scoville2007}, using 600 hours of observations
with the VIMOS spectrograph at VLT. It consists of two parts:
zCOSMOS-bright and zCOSMOS-deep. zCOSMOS-bright is a survey purely
magnitude limited in I-band; when complete it will provide a sample of
$\sim 20000$ galaxies in the range $15.0 \leq I_{AB} \leq 22.5$ from
the HST ACS imaging \citep{Koekemoer2007} over the whole area of 1.7
deg$^2$ of the COSMOS field. zCOSMOS-deep targets $\sim 10000$
galaxies, selected through color criteria to have $1.4 \la z \la 3.0$
within the central 1 deg$^2$. At completion it will provide redshifts
for $\sim 10000$ galaxies with magnitudes $B_{AB} \leq 25.0$.

\begin{figure*}
\centering                  
\includegraphics[width=17cm,angle=0]{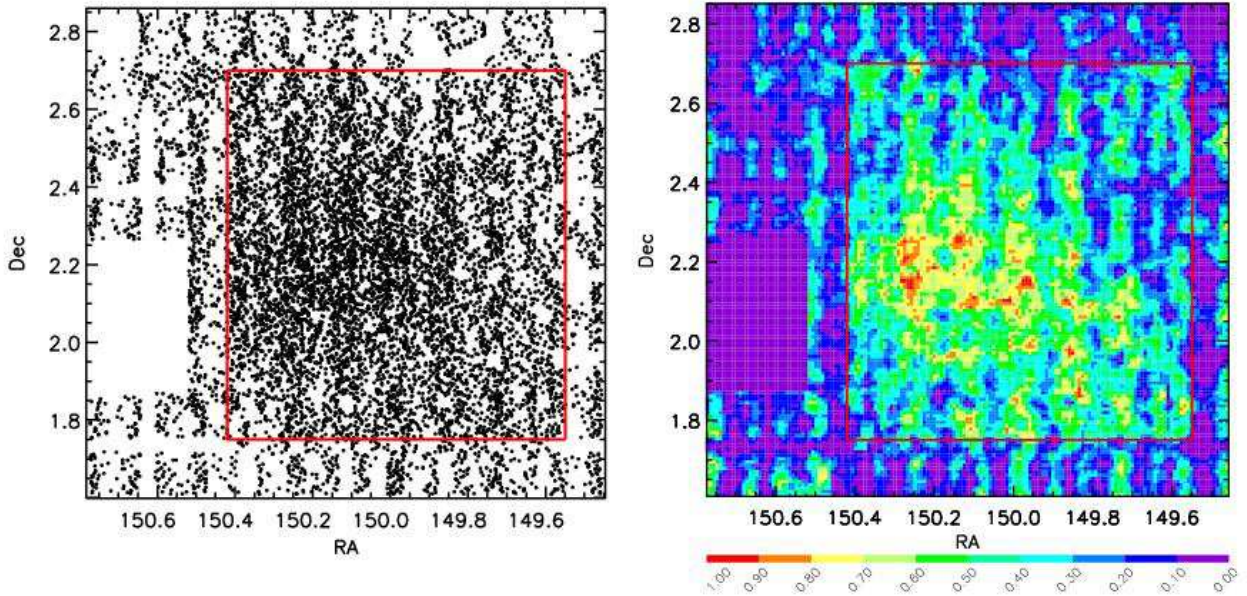}  
\caption{The left panel shows the $ra-dec$ distribution of the $\sim
10000$ objects observed in the first half of the zCOSMOS-bright
survey.  The right panel shows the $ra-dec$ distribution of the ratio
of the number of objects with reliable spectroscopic redshift to the
total number of potential targets \ie non-stellar objects in the
parent bright photometric catalogue ($15.0 \leq I_{AB} \leq
22.5$). The color scale provides the legenda for the range of values
displayed on the plot. In both panels the red rectangle indicates the
restricted area used for the analysis presented in this paper,
corresponding to $\sim 0.83 sqdegs$. Within this restricted area the
mean value of the sampling rate is around $40$\%, \ie two objects out
of five have a reliable redshift measured (see text for more
details). }
\label{fig:radec_distr_map}
\end{figure*} 

The analysis presented in this paper uses of the sample of $10644$
objects for which we obtained spectra during the first half of the
zCOSMOS-bright observational campaign.  This number corresponds to a
total of 83 pointings of the VIMOS spectrograph, observed during ESO
periods P75, P76, and P77 and includes compulsory, \ie objects with
forced slit positioning, and secondary targets, \ie objects
serendipitously falling inside the slit other than the primary target.

zCOSMOS-bright observations use the $R = 600$ MR grism and 1 hour
integrations to secure redshifts with a high success rate. The
wavelength range covered goes from $5500 \leq \lambda \leq 9700$
\AA. For more details of the survey strategy and characteristics we
refer the reader to \citet{Lilly2007} and Lilly et al. (2009).
 
The distribution on the sky of the $\sim 10000$ objects observed in
the first half of the zCOSMOS-bright survey is illustrated in the left
panel of Fig.~\ref{fig:radec_distr_map}.  The vertical banding
visible in the external, less finely sampled, regions reflects the
quadrant design of VIMOS and an additional pattern introduced by the
slit positioning software SPOC \citep{Bottini2005}.  This pattern
should almost completely disappear at survey completion, since the
observational strategy foresees an eight-pass coverage with two mask
designs at each pointing.  The expected final sampling rate is around
$\sim 70$\%.

The redshift distribution of the observed galaxies covers the range
$0.1 \leq z \leq 1.2 $ and peaks at redshift $\sim 0.7$.  The error in
the redshift measurement, as determined from repeated observations, is
around 100 \kmsec, an accuracy well suited to the original survey
scientific goals of the investigation of large scale structure and the
detection of groups.

For each measured redshift, we adopted a ranking scheme reflecting our
confidence in its correctness.  It is based on six broad confidence
classes (0-1-2-3-4-9) reflecting the quality of the redshift
measurement as obtained from the spectra. This scheme is similar to
that originally adopted in the CFRS \citep{LeFevre1995} and VVDS
\citep{LeFevre2005}, but with some further refinements taking
advantage of the wealth of photometric information available for each
targeted object.

The large, exquisite quality, ancillary photometric database provided
by the COSMOS survey (from HST data, to Spitzer, Galex, Chandra,
CFHTLS, and Subaru data, see \citet{Scoville2007}) has enabled us to
derive reliable photometric redshifts for all objects in the
zCOSMOS-bright parent photometric catalogue, with an uncertainty as
low as $\Delta z \sim 0.01 \times (1+z)$ \citep{Ilbert2009}.

The photometric redshift information was used to incorporate in our
analysis objects whose spectroscopic redshift, although less secure,
was consistent with its photometric value and therefore deemed
reliable ($\delta z$ smaller that ~$0.08 \times (1+z)$, see Lilly et
al., 2009, for more details). In this way one can use $\sim 85$\% of
the observed sample, totalling $\sim 8600$ galaxies up to $z=2.0$
($\sim 9200$ including stars and with no high redshift cut-off), with
a nominal spectroscopic confirmation rate of $\sim 98.5$\% as found by
duplicate observations. This sample represents roughly half of the
final zCOSMOS sample and when we talk of the \10K we always refer to
this subset of objects, the same used to perform group searches
\citep{Knobel2009}.

For the \10K galaxies, absolute rest-frame magnitudes and stellar
masses were obtained using standard multicolor spectral energy
distribution (SED) fitting analysis.  Rest-frame absolute magnitudes
were obtained using the ZEBRA code, for which a detailed description
is provided in \citet{Feldmann2006} and Oesch et al. (in prep.). We
note here that the templates used by ZEBRA are the standard CWW
templates \citep{Coleman1980} and starburst templates from
\citet{Kinney1996}, and the best fit template is normalized to each
galaxy photometry and spectroscopic redshift.

Stellar masses in units of solar masses were obtained by fitting
stellar population synthesis models to the multicolor spectral SED of
the observed magnitudes using the {\it Hyperzmass} code
\citep{Bolzonella2009, Pozzetti2009}. In the subsequent analysis, we
use stellar masses calculated adopting the \citet{BruzualCharlot2003}
libraries, and assuming a Chabrier initial mass function
\citep{Chabrier2003}. More details about the {\it Hyperzmass} code can
be found in \citet{Bolzonella2009}.

\begin{figure*}
\centering                  
\includegraphics[width=17cm,angle=0]{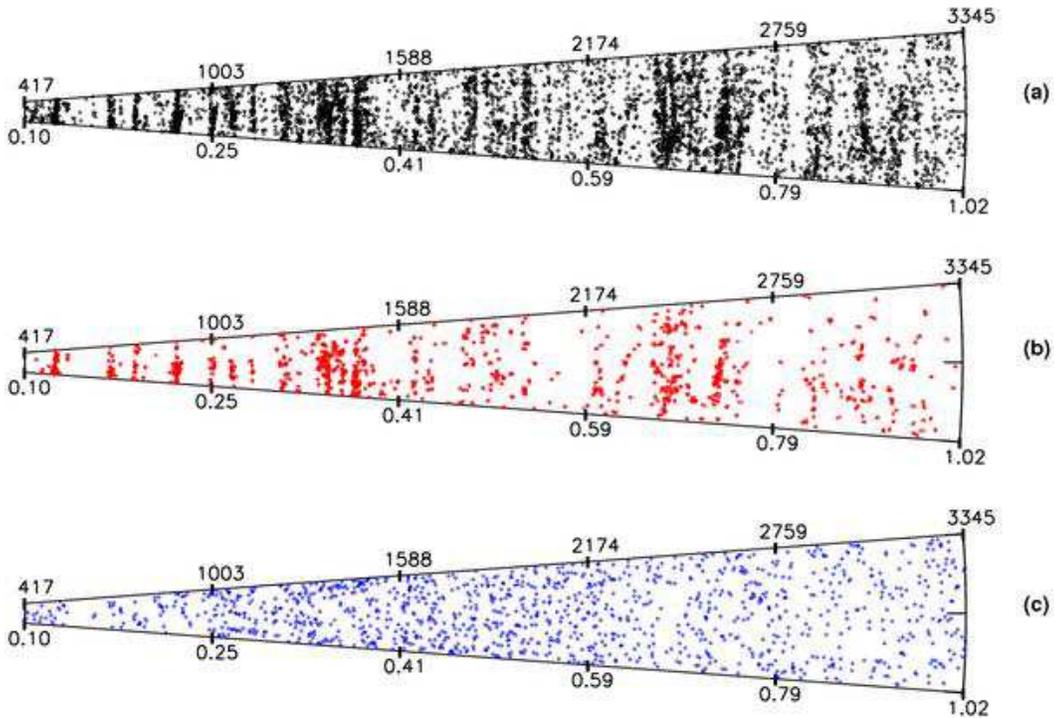} 
\caption{Panel (a) shows the distribution along the line of sight of
the 6204 galaxies used in our analysis, the subset of the \10K
galaxies sample located within the central, high sampling rate, region
of the survey and in the redshift range $ 0.1 \leq z \leq 1.0 $.
Panel (b) shows the distribution along the line of sight of the 1966
group galaxies, while panel (c) show the same for the sample of 1146
isolated galaxies. In each panel we collapsed the $dec$ axis and
expanded by a factor of $\times 10$ the distances corresponding to the
$ra$ axis to make the plot easier to read. Labels on the top of the
cone indicate the line-of-sight distance in units of comoving \Mpch,
while those on the bottom indicates the redshift. The transverse
dimension at redshift $z=0.1$ is $\sim 7$ \Mpch, while that at
redshift $z=1.0$ is $\sim 50$ \Mpch, corresponding to the dimension
of $\Delta ra \sim 0.87 ~deg$ of our restricted sample.}
\label{fig:all_cones}
\end{figure*} 

Panel (a) of Fig.~\ref{fig:all_cones} shows the distribution along
the line of sight of the galaxies within the boundaries defined by
$149.55 \leq ra \leq 150.42 $, $ 1.75 \leq dec \leq 2.70 $ and $ 0.1
\leq z \leq 1.0 $. These boundaries (see
Section~\ref{sect:IncomplCorr}) are those adopted in our analysis to
avoid being affected by the inhomogeneous coverage of the \10K. The
number of galaxies surviving within these $ra-dec-z$ boundaries equals
6204, of which 1966 are in groups, while 1146 define the so-called
isolated galaxy sample (see Section~\ref{sect:sample10Kisolated}).

The transverse dimension of this restricted sample along $ra$ is $\sim
7$ \Mpch at $z=0.1$, and $\sim 50$ \Mpch at $z=1.0$. In the same
redshift interval the total contiguous comoving volume sampled is $
\sim 3.3 \times 10^6$\Mpch$^3$.

\subsection{The 10K group catalogue}
\label{sect:sample10Kgroups}  

In our analysis, we use the catalogue of groups presented in
\citet{Knobel2009} and refer the reader to that paper for a detailed
presentation of both the group finding algorithm and the group
catalogue. Here we summarize the main points and advantages of the
adopted group finding algorithm and briefly discuss the resulting
group catalogue.  \citet{Knobel2009} introduced a novel method,
defined as a ``multi-pass procedure'', to achieve an impressive
quality in group reconstruction as tested using realistic mock
catalogues. This method, when combined with the standard
fried-of-friends (FOF) algorithm, yields values of completeness and
purity for the group catalogue obtained that are extremely stable with
both redshift and number of members observed in the reconstructed
groups. Typical values of these two quantities for groups
reconstructed with more than five observed members are around $\sim
80$\% at all redshifts and do not decrease substantially for groups
with lower number of members. Correspondingly the interloper fraction
always remains below $\sim 20$\% at all redshifts for groups
reconstructed with more than five observed members, with only a slight
increase for groups with lower number of members \citep{Knobel2009}.

These results provide reassurance that the group catalogue that we use
in our subsequent analysis is highly homogeneous up to $z \sim 1$, a
fundamental prerequisite, since the aim of this paper is to explore
redshift trends in group galaxy colors.  If our results are to be
reliable, we need to be confident that the group catalogue we use is
almost entirely free from redshift dependent biases. The presence of a
significantly higher interloper fraction with increasing redshift
could surreptitiously increase the fraction of blue (field) galaxies
observed in our group catalogue and be mistakenly interpreted as
evidence of evolution.  The extensive tests performed in
\citet{Knobel2009} place on solid basis the analysis that we perform
in the following sections.

Panel (b) of Fig.~\ref{fig:all_cones} shows the distribution along
the line of sight of the group galaxy population (1966 galaxies in
total) within the boundaries defined by $149.55 \leq ra \leq 150.42 $,
$ 1.75 \leq dec \leq 2.70 $ and $ 0.1 \leq z \leq 1.0 $.  The presence
of large structures is clearly delineated by the group galaxy sample.
In fact there are quite a few conspicuous structures visible in this
plot, \eg those around redshifts $\sim 0.35$ and $\sim 0.7$, while
there are, on the other hand, regions devoid of large structures, \eg
in the redshift range $ 0.4 \leq z \leq 0.6$ \citep[see][for a
detailed description of the density field structures in the \10K
field]{Kovac2009a}.

We also note that our survey does not contain any single rich cluster,
for example comparable to Coma cluster in the local Universe. This is
not unexpected: because the size of the volume of the Universe
explored by zCOSMOS-bright the probability of one such cluster being
observed is negligible \citep[see also][]{Finoguenov2007}.    

\subsection{The isolated galaxy sample}
\label{sect:sample10Kisolated}
  
We complemented the analysis performed on the sample of group galaxies
with a parallel one on a sample of isolated galaxies, \ie a sample of
galaxies located in low density regions. This comparative analysis
should highlight the differences -- if any -- in properties (rest
frame colors in our analysis) from the group galaxy sample and
therefore allow us to quantify the environmental dependencies of the
properties explored more reliably.

To define the isolated galaxy sample, we use the Voronoi Tessellation
method \citep{Voronoi1908}. Voronoi Tessellation divides the space
occupied by the survey into a set of unique polyhedral sub-volumes,
each containing exactly one galaxy and all points in space that are
closer to that galaxy than to any other. As a consequence, while
galaxies with many neighbors (\eg those in groups and high density
environments) have small Voronoi volumes, relatively isolated galaxies
have larger Voronoi volumes. Voronoi Tessellation has been used in the
literature as a basis for group-finding algorithms
\citep{Marinoni2002, Gerke2005, Cucciati2009a, Knobel2009}. It is
quite straightforward to use Voronoi volumes to select a sample of
isolated galaxies, defined as galaxies occupying the largest Voronoi
volumes. This strategy has the advantage of being non-parametric, \ie
it avoids any arbitrarily chosen smoothing/window profile in defining
low density regions.

However, proper attention must be taken to exclude galaxies that are
close to survey borders and correct for the progressive increase in
the typical size of Voronoi volumes between low and high redshifts
in our flux-limited galaxy sample.

To avoid the first problem, \ie of galaxies near the survey boundaries
entering the isolated galaxy sample because of their apparently large
Voronoi volumes, we decided to restrict the volume of the search for
isolated galaxies within the boundaries defined by $149.57 \leq ra
\leq 150.41 $, $ 1.76 \leq dec \leq 2.68 $ and $ 0.1 \leq z \leq 1.0
$, which is slightly more restrictive than the limits adopted for the
group analysis indicated by the red lines in
Fig.~\ref{fig:radec_distr_map}. Furthermore, in all the subsequent
analysis we decided to reject all isolated galaxies located in areas
of lower sampling, that is galaxies with mean correction factor
$~\psi(\alpha,\delta) \ge 5$ (see Section~\ref{sect:IncomplCorr}). For
these galaxies, a large measured Voronoi volume could be the result of
the low spectroscopic sampling rate in the surrounding area.

\begin{figure} 
\includegraphics[width=9cm,angle=0]{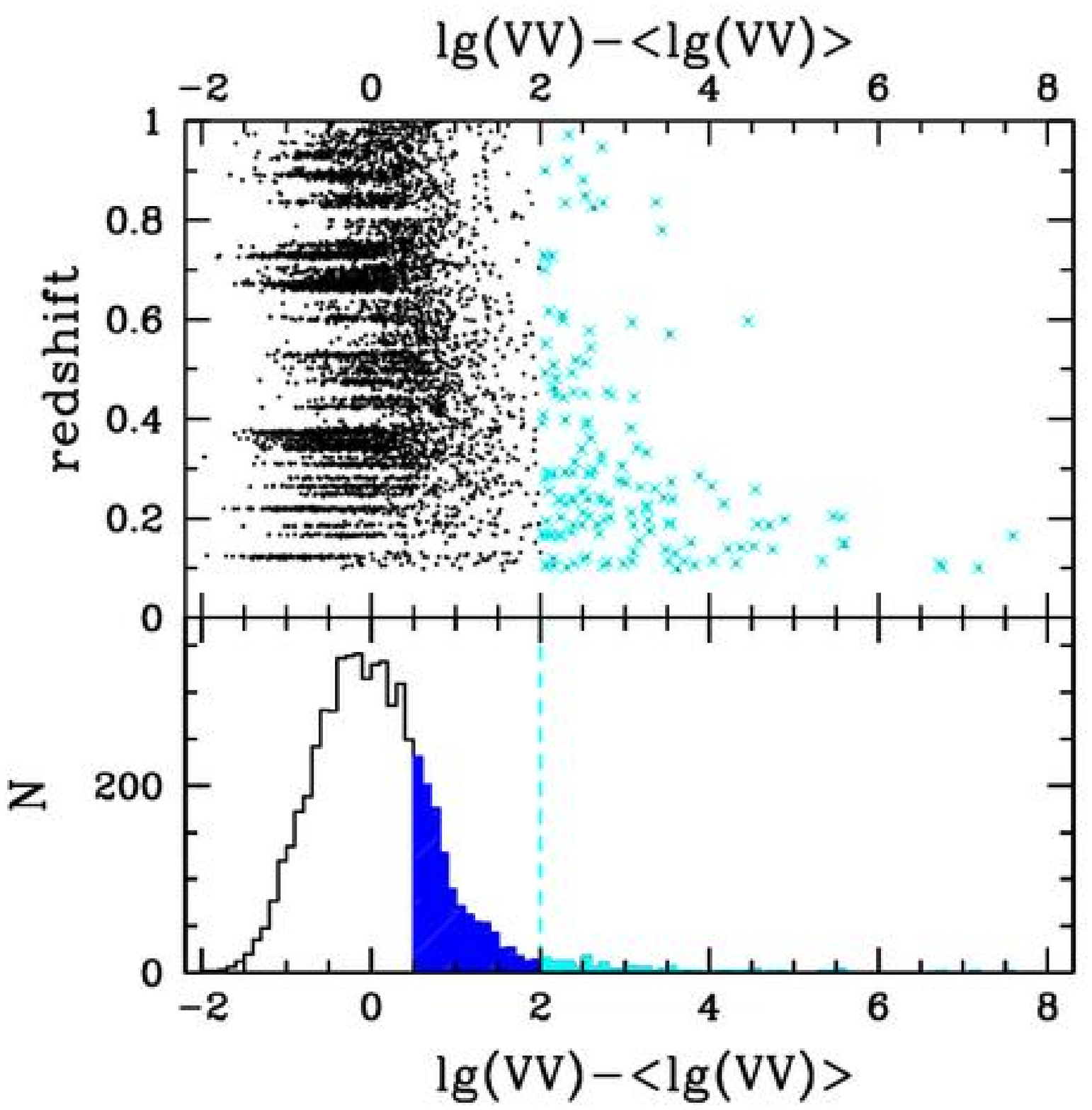} 
\caption{Top panel: Distribution of logarithm of normalized Voronoi
volumes as a function of redshift. The stripes extending towards lower
Voronoi volumes values are due to the presence of groups. Points
marked with a cyan cross correspond to galaxies removed from the
isolated galaxy sample because their Voronoi volume exceeded by a
factor of $100$ that of the  median at the corresponding
redshift. Bottom panel: histogram of the total distribution of
normalized Voronoi volumes. The shaded blue area corresponds to the
last quartile of the distribution, chosen to select our isolated
galaxies sample, while the long tail in cyan extending to higher
values - and indicated also by the vertical line - is the one
clipped from the sample.}
\label{fig:isol_selection}
\end{figure} 

To avoid the second possible problem, of being biased in the
definition of isolated galaxies by the progressive decrease in the
galaxy density in our flux-limited sample, we computed the median
value of the logarithm of Voronoi volumes sizes as a function of
redshift using running bins of size $\Delta z \leq 0.2$ in redshift
steps of $0.05$.  A simple linear fit to this quantity (as deemed
reasonable by visual inspection) was then used to normalize all
measured Voronoi volumes, correcting for the progressive increase with
redshift in the mean inter-galaxy separation. We then selected the
highest quartile of the normalized volumes distribution obtained in
this way, after taking the simple precaution of further rejecting
galaxies (148 in total, 80 at $ z \leq 0.25$) whose normalized Voronoi
volume was more than 100 times larger than the median one: mostly
galaxies located too close to the survey borders, as suggested by the
large predominance of low redshift objects and by their general
distribution on the sky.  Figure~\ref{fig:isol_selection} illustrates
the method adopted to select isolated galaxies. The final number of
isolated galaxies obtained this way is 1146, after the removal of
galaxies (206, out of which 128 located in pairs) listed in our group
catalogue.

We checked the reliability of our approach by selecting isolated
galaxies in simulations. We used the 24 COSMOS mock light-cones kindly
provided by M. Kitzbichler, \citep{KitzbichlerWhite2007}, based on the
Millennium DM N-body simulations \citep {Springel2005}. We applied the
same observational strategy to these cones used to select the \10K: we
chose the same pointings observed in \10K, used SPOC to select {\it
observed} targets and included the same redshift success rate as
the real data.  Out of the sample of isolated galaxies obtained from
the mocks using the procedure described above, $\sim 60$\% are truly
isolated galaxies,  with a variance of a few percent from cone to
cone, \ie galaxies that in mock light-cones are inside a halo that
contains only one galaxy down to the $R=26$ magnitude
limit. However, when considering the 10K mock samples - limited
to $I_{AB} = 22.5$ and with our sampling rate applied - this number
increases to $\sim 90$\%. In other words, only $\sim 10$\% of the
galaxies in the isolated sample selected using our strategy are
located in groups with at least two members in the 10K mock samples.
We should get rid of most of this contamination by the last step of
our procedure: the final trimming of galaxies listed in our real group
catalogue.

Panel (c) of Fig.~\ref{fig:all_cones} shows the distribution along the
line of sight of the isolated galaxy sample, whose uniformity is
evident.

\section{Measuring \bluefrac}
\label{sect:sampleanalysis}  

We use in this paper the diagnostic tool introduced in the literature
in the seminal work by \citet{ButcherOemler1978}.  These authors were
the first to note that the fraction of blue galaxies (\bluefrac from
now onwards) in clusters seems to increase with redshift. Their work
started a long-lasting wave of observational and theoretical papers,
which is still far from being completed (see the short literature
review presented in the introduction).

After thirty years, the value of \bluefrac is still a valuable and
effective empirical tool in studying of the dependence of
galaxy evolution from the environment in which they reside.

Galaxy color is the easiest parameter to measure among those that
exhibit a distinctive bi--modality: spectral class, morphology, star
formation rates and metallicities \citep[see][]{Strateva2001,
Mignoli2009}.  Therefore it is the simplest to adopt in parametrizing
the differences between evolution of groups and field or isolated
galaxies.  As far as its physical meaning is concerned, the rest-frame
$(U-B)$ color adopted in our analysis, bracketing the 4000\AA ~break,
can be used to study the average star formation histories over longer
time-scales than emission lines indicators such as, \eg [OII].  This
choice could therefore provide clearer insights into mechanisms that
operate on longer time scales, such as possibly those in action in
dense environments as groups, where member galaxies have resided for a
significant fraction of their lifetime. 

Despite the apparent simplicity of this parameter, the origin of the
physical mechanisms responsible for the variations in \bluefrac in the
group/cluster population still remains to be fully explained. In
particular we are still unable to determine the relative influences of
processes related to the environment and those that are intrinsic to
the galaxy itself, and therefore the dichotomy between
ab-initio/internal and external mechanisms responsible for the
variation of \bluefrac is still an open one.  The fraction of galaxies
on either sides of the bimodality in $(U-B)$ colors has been shown to
depend strongly on galaxy luminosity and stellar mass \citep[see,
\eg][]{Baldry2004, Baldry2006}. Therefore, in studying the dependence
of \bluefrac on group environment, we define and adopt both luminosity
volume-limited and mass volume limited samples.

In this Section, we discuss the strategy adopted to correct for the
\10K incompleteness when measuring $\it{F_{blue}}$, the cut-off
adopted in defining $\it{F_{blue}}$, and how we estimate errors on
this quantity.

\subsection{Correcting for survey incompleteness}
\label{sect:IncomplCorr}   

The left panel of Fig.~\ref{fig:radec_distr_map} shows that the
coverage in $ra-dec$ of the \10K remains very uneven.  While the mean
sampling rate of the \10K is around $\sim 30$\%, this number varies
significantly as a function of position: in the central regions the
sampling rate is as high as $\sim 70$\%, while it is as low as $\sim
10$\% in the regions near the borders. This unevenness can create
problems when defining groups of homogeneous numerosity/richness
irrespective of their position in the sky (see Section
~\ref{sect:GroupRich}).

To correct for this problem we adopted for each galaxy a weighting
scheme consisting of two factors: $~\phi(m)$ and
$~\psi(\alpha,\delta)$. The first factor $\phi(m)$ is similar to one
adopted for the luminosity and mass function estimates (see Zucca et
al., 2009, for more details). It is obtained by a parabolic fit to the
product {\it W} of the inverse of the target sampling rate ($TSR$) and
the inverse of the spectroscopic sampling rate ($SSR$):

\begin{equation}
         {\it W } =(1/TSR)*(1/SSR)
\end{equation}

\noindent $TSR$ is defined as ${TSR} = N_{obs}/N_{phot}$, the ratio of
the total number of objects observed to the total number of 
potential targets, \ie non-stellar objects in the parent bright
photometric catalogue, \citep[see][]{Lilly2009}. For the few
compulsory targets observed in our survey (\ie with forced slit
positioning) TSR was defined to equal 1. $SSR$ is defined as ${SSR(m)}
= N_{spec}(m)/N_{obs}(m)$ the ratio, calculated in bins of apparent
magnitude, of the number of observed objects whose redshift was
reliably measured to the total number of observed objects. The
apparent magnitude dependence takes into account the progressive
difficulty, moving toward fainter magnitudes, to measure a redshift. A
more complex scheme, which includes the redshift dependence of SSR
does not alter appreciably the final results (see Bolzonella et al.,
2009).

The second factor $~\psi(\alpha,\delta)$ corrects for the variation,
as a function of $ra-dec$, of the mean correction factor expressed by
$~{\phi}(m)$. We estimated $~\psi(\alpha,\delta)$ in two passes. In a
grid of steps equal to $30\arcsec$ in right ascension and declination
and in squares of $2\arcmin\times2\arcmin$, we computed the ratio of
the number of observed objects whose redshift was reliably measured to
the total number of potential targets,  as defined as above,
within the same area. We then obtained $~\psi(\alpha,\delta)$ by
normalizing to unity the mean value of this ratio over the full
$ra-dec$ coverage of the \10K survey. The right panel of
Fig.~\ref{fig:radec_distr_map} shows in color-scale the resulting
function $~\psi(\alpha,\delta)$ before normalization. The parameters
chosen in calculating this function allow us to reproduce well the
inhomogeneities in the survey, even the vertical banding, visible in
left panel of Fig.~\ref{fig:radec_distr_map}.  To each galaxy we
therefore assigned a weight: $w_{i} = \phi(m)\times
~\psi(\alpha_i,\delta_i)$ which is the galaxy weighting scheme used in
the following analysis.

At the borders of the survey sampling is lower than average resulting
both in higher galaxy weights and higher incompleteness in group
detection. To alleviate this problem, we decided to restrict the
analysis to the central area of the survey, where the inhomogeneity in
sampling rate is significantly lower. This region is indicated by
red lines in Fig.~\ref{fig:radec_distr_map} and corresponds to
galaxies within the following boundaries: $149.55 \leq ra \leq 150.42
$, $ 1.75 \leq dec \leq 2.70$.

We note that our results are relatively insensitive to changes in the
strategy used to define the weights, for example larger smoothing
boxes in defining $\psi(\alpha_i,\delta_i)$.  Even when no weights at
all are used, our results are almost unchanged.  A weighting scheme is
needed when estimating in a homogeneous way group richness (for
example when exploring trends of \bluefrac as a function of groups
richness). When dealing with the galaxy group population as a whole,
the impact of the use of weights is minimal.

\subsection{Computing the blue fraction}
\label{sect:CompBlueFrac}   

We divided galaxies into red and blue sub-samples taking advantage of
the observed bimodality in galaxy $(U-B)$ rest-frame colors, visible
in Fig.~\ref{fig:colmag} (see also Cucciati et al.,
2009). Accordingly, we defined blue galaxies as those with rest-frame
colors $(U-B) \leq 1.0 $. This value agrees with both the value chosen
by \citet{Gerke2007} in their analysis of \bluefrac in the DEEP2
groups sample and with the value adopted in a parallel analysis to our
own by \citet{Cucciati2009a}. We did not allow this value to vary with
galaxy luminosity, as suggested for example by \citet{vanDokkum2000}
and \citet{Blanton2006a}.  Given the relatively small variation in
$M_B$ of the bulk of our galaxy sample (of roughly 3 magnitudes), the
color-magnitude relationship quoted by these authors would imply a
corresponding variation in the cut-off color value $\leq 0.1$ mags,
which we deemed to be negligible.

From our data, there is no obvious evidence of evolution to redshift
$\sim 1$ in the adopted cut-off value, and in our analysis we
therefore decided to keep its value fixed with redshift.

After defining the cut-off value between red and blue galaxies we
obtained a set of ${N}_{b}$ blue galaxies from the total sample of
${N}_t$ galaxies, each with a weight $w_i$.  The corrected blue
fraction was then given by:

\begin{equation}
     \it{F_{blue}} = {\cal N}_{b} / {\cal N}_{t}
\end{equation} 

\noindent where the number of blue galaxies ${\cal N}_{b}$ and the total number
of galaxies ${\cal N}_{t}$ are defined to be:

\begin{equation}
     {\cal N}_b = \sum_{j=1}^M w_j  ,  ~~ {\cal N}_t = \sum_{i=1}^N w_i 
\end{equation} 

\noindent where the index $j$ corresponds to all the blue galaxies,
while the index $i$ corresponds to the full galaxy sample.

\subsection{Estimating errors in \bluefrac}
\label{sect:ErrBlueFrac}   

To estimate errors in the values computed for $\it{F_{blue}}$, we adopted a
bootstrap re-sampling strategy. We randomly sampled by replacement the
entire data set under consideration, \eg all isolated galaxies in a
given volume-limited sample. The error in \bluefrac was then
estimated to be the standard deviation in \bluefrac distribution
for 1000 such Montecarlo samples.

We used also the approximate analytical formulas provided by
\citet{Gehrels1986} to estimate the error in \bluefrac but the
differences in value with respect to the bootstrapping technique are
minimal. In our plots we always show bootstrap errors.

Another source of errors and noise in our plots is cosmic variance.
At lower redshifts, the volume sampled by zCOSMOS survey is not large
enough to be considered a fair representation of the universal matter
distribution. It is therefore possible that the presence of large
scale structures introduces large fluctuations in the trends of
\bluefrac as a function of redshift, lowering significantly \bluefrac
at the redshift where these structures are located. Our survey shows
quite a few of these prominent structures, for example those located
at $z \sim 0.35$ and $z \sim 0.7$, readily visible in the top two
panels of Fig.~\ref{fig:all_cones} (see also Kovac et al., 2009).  To
alleviate this problem in our analysis we tried to adopt redshift bins
large enough to smooth out as much as possible this effect.

\section{Defining luminosity volume-limited samples}
\label{sect:VolLim}  

The zCOSMOS survey provides a unique data-set for measuring the
evolution of the blue fraction up to $z \sim 1$.   The excellent
quality of the observed spectra prevent any possible bias against red,
absorption lines only spectra \citep{Lilly2009}, while the simple
$I_{AB} \leq 22.5$ magnitude limit used to select survey targets
translates into a selection in the rest-frame $B$-band at $z \sim
0.8$. Therefore the zCOSMOS galaxy sample when rest-frame B-band
selection is adopted is free from significant color-dependent
incompleteness in $(U-B)$ rest-frame colors to the highest redshift
bin explored.

However the reader should be warned that completeness in B-band
rest frame selection does not imply completeness in, \eg mass
selection, as we will discuss at lenght in Section~\ref{sect:RedefBO}
and following.  As a consequence any trend observed in rest-frame
B-band selected samples needs to be re-examined when the selection
criterion of the sample changes \citep[see also, \eg][]{DePropris2004}.

\begin{figure}
\includegraphics[width=9cm,angle=0]{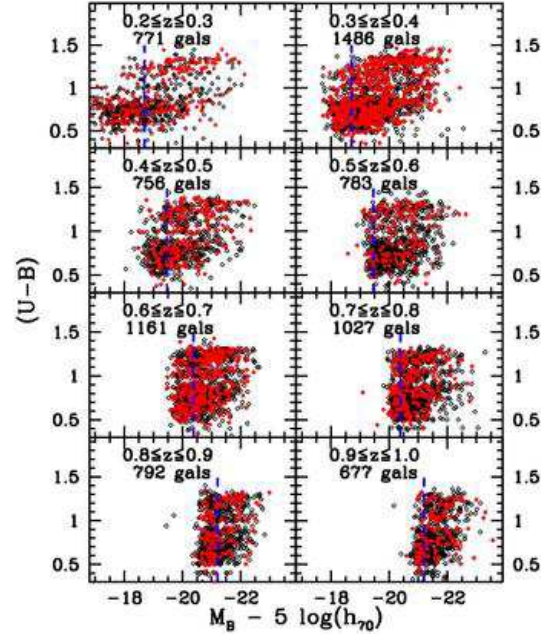} 
\caption{Rest-frame (U-B) colors plotted vs $B-$band rest-frame
magnitudes. In each panel redshift bins of width $\Delta z = 0.1$ have
been considered, as indicated by the labels, that list also the total
number of galaxies located in each redshift bin. The points in red
show galaxies located in groups according to the catalogue obtained
from the full \10K. The vertical dashed line in each panel indicates
the absolute magnitude limits corresponding to the volume-limited
sample that was chosen for galaxies contained in each redshift range.}
\label{fig:colmag}
\end{figure}

The absence of $(U-B)$ color incompleteness in zCOSMOS B-band
volume limited samples can be visually appreciated in
Fig.~\ref{fig:colmag}, where we plot for different redshift bins (as
indicated in each panel) the rest-frame $(U-B)$ color \vs rest-frame
$B-$band absolute magnitude $M_{B}$.

In each panel distinctive red and blue populations of galaxies are
visible, with loci that are separated approximately at $(U-B) = 1 $ at
all redshifts.  The cut-off in the galaxy population distribution
visible on the left hand side of each panel is a consequence of
zCOSMOS purely $I-$band flux-limited target-selection strategy and
moves towards brighter magnitudes as the redshift increases. However,
this progressively brighter cut-off does not introduce obvious biases
against red galaxies as indicated by the cut-off line being nearly
vertical in all panels with the possible exception of the last
redshift bin, where the observed $I-$band begins moving blue-ward of
the rest-frame $B-$band and a slight slanting of the cut-off in the
galaxy population distribution starts becoming appreciable. Therefore,
in this last redshift bin we need to be more conservative in the
definition of the cut-off in absolute $B$-band rest-frame magnitude to
avoid biases against red galaxies, even if this choice will result in
smaller number of objects for our analysis.

Another factor to consider in our definition of volume-limited samples
is that the typical galaxy luminosity evolves with redshift.  We need
to include an evolutionary term in our definition of cut-off
magnitudes for the volume-limited samples because we aim to select a
population of galaxies that is similar with respect to $M^*_{B}$ at
all redshifts.

As suggested by the results obtained for the global luminosity
function evolution of our sample (see Zucca et al., 2009, for more
details), the evolution in $M^*_{B}$ can be parametrized linearly by
the equation

\begin{equation}
M^*_{B \it ev} = -20.3 - 5\times log~(h_{70}) -1.1\times z,
\end{equation} 

\noindent which includes an evolution with redshift of roughly 1
magnitude between $z\sim 0.1$ and $z\sim 1$ for $M^*_{B}$.

The galaxy luminosities quoted from now on are always
evolutionary--corrected present--day luminosities to ensure that
galaxies of similar luminosity are being compared in different
redshifts bins. 

\begin{figure}
\includegraphics[width=9cm,angle=0]{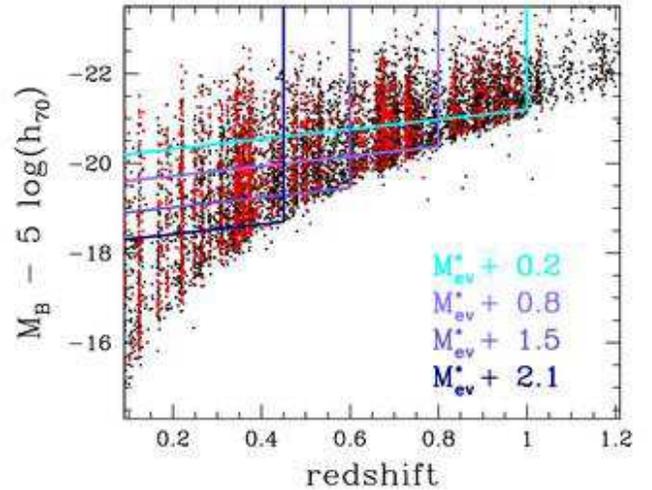}
\caption{Redshift distribution of the \10K zCOSMOS galaxies. Red
points represent galaxies located in groups. The lines drawn
correspond to the four different volume-limited samples discussed in
the text. We assumed $M^*_{\it ev} = -20.3 - 5\times log~(h_{70})
-1.1\times z$ and the different labels and the lines drawn correspond
to the four different volume-limited samples discussed in the text. }
\label{fig:span_lum}
\end{figure}

We defined four different luminosity volume-limited samples, from
sample I to sample IV, each covering progressively higher ranges of
redshift, and defined by evolving the cut-off magnitudes $M_{cut-off}
= M^*_{B \it ev} + 2.1/+1.5/+0.8/+0.2 $, as illustrated in
Fig.~\ref{fig:span_lum}.

Table~\ref{tab:vollimnumb} summarizes the properties of these four
different volume-limited samples: the different redshift ranges
covered and the total numbers of galaxies and isolated/group galaxies
contained within the $ra-dec$ limits described in
Section~\ref{sect:IncomplCorr}.

From now onwards when we will talk of the field population we always
mean the total galaxy sample, \ie the full galaxy population including
group/isolated galaxies.

We note that while the full group catalogue was obtained using the
entire \10K galaxy catalogue, for each of the volume-limited samples
defined above we selected a corresponding uniform sample of groups
possessing at least two member galaxies brighter than the B-band
rest-frame $M_{cut-off}$ considered (Group galaxies I). This strategy
avoids the redshift inhomogeneity introduced in our group catalogue by
the progressive brightening of the rest-frame $B-$band magnitudes
sampled by the survey as redshift increases. A given group will have a
different number of members in each volume-limited sample, but within
each volume-limited sample group's numerosity/richness will be
measured consistently at all redshifts. Unless explicitly mentioned
when we talk of group galaxies, we refer to Group galaxies I.

We also introduced a further set of galaxy groups: those that possess
at least two members in sample IV (Group galaxies II). By studying the
variation in \bluefrac for galaxies of different luminosities that are
members of this group sample one can hope to disentangle the effect of
galaxy luminosity on \bluefrac from that of group richness: this is
because the groups in this sample should be homogeneous in terms of
richness as a function of redshift, irrespective of the magnitude of
the member galaxies considered in the analysis (see
Section~\ref{sect:VolLimSamples}). For the sake of robustness, the
value of the group observed line-of-sight velocity dispersion
$\sigma$, whenever used in our analysis, is always estimated using all
observed group members, irrespective of their absolute magnitude.

\begin{table*}
\caption{Summary of the four volume-limited data samples. 
We assume $M^*_{\it ev} = -20.3 - 5~log~h_{70} -1.1~z$.} 
\label{tab:vollimnumb}  
\centering                  
\begin{tabular}{c c c c c }    
\hline\hline                 
                & Sample I  & Sample II & Sample III & Sample IV  \\   
 $M_{B}$ range  & $M_{B}\leq M_{\it ev}^*+2.1$ & $M_{B}\leq M_{\it ev}^*+1.5$ & $M_{B}\leq M_{\it ev}^*+0.8$ & $M_{B}\leq M^*_{\it ev}+0.2$ \\
    $z$ range   & $0.1\leq z\leq0.45$ & $0.1\leq z\leq0.6$ & $0.1\leq z\leq0.8$ & $0.1\leq z\leq 1.0$ \\                  
\hline \hline                    
All galaxies       &     1798   &  2122     &  2616    &  2182  \\      
Isolated galaxies  &     315    &   442     &   431    &   326  \\
Group galaxies I   &     676    &   670     &   709    &   447  \\
Group galaxies II  &     218    &   237     &   412    &   447  \\
\hline                             
\end{tabular}
\end{table*}

After defining galaxy weights, volume-limited samples and the
corresponding group/isolated subsets, we proceeded to estimate
$\it{F_{blue}}$, the fraction of blue galaxies, for each galaxy sample
and its dependence on group properties, galaxy luminosity, and
redshift.

\section{Blue fraction as a function of galaxy luminosity and environment up to $z \sim 1$}  
\label{sect:VolLimSamples} 

In the local Universe, the correlation between galaxy luminosities and
colors is a well-known observational result: more luminous galaxies
have typically redder colors than less luminous galaxies
\citep[see][and references therein]{Baldry2004}. A similar color
segregation has been observed between local groups and field samples:
redder galaxies are preferentially located in galaxy groups and
clusters \citep[see][and references therein]{DePropris2004}.  It is
therefore interesting to use our sample to check whether these trends
survive at higher redshifts and if they show weakening or even any
visible reversal.

A similar analysis was performed using DEEP2 data for the redshift
range $0.75 \leq z \leq 1.3$ by \citet{Gerke2007}, and using VVDS data
for the range $0.25 \leq z \leq 1.2$ by \citet{Cucciati2009a}.  The
VVDS and DEEP2 surveys were the first to use in their investigation a
homogeneous dataset from the lowest to the highest redshift bins
explored, and a group sample spanning a wide range of richnesses, down
to poorest systems, in contrast to previous work that mainly 
considered higher richness, and more easily detectable, systems.  With
respect to these two pioneering large high-redshift surveys, zCOSMOS
presents some non-negligible advantages. We have a larger volume
coverage enabling us to complete more robust statistical analyses than
VVDS, and smaller errors in galaxy redshift measurements  - around
275 \kmsec for VVDS, \citep[see][]{LeFevre2005}, which allows us to
compile a group catalogue that is less prone to contamination and
incompleteness, especially for low richness and low velocity
dispersion systems. We are also less plagued by the color
incompleteness (and, more importantly for the subsequent analysis,
mass incompleteness) that affects DEEP2 data in the redshift range
covered by their analysis, and have the ability to cover the complete
redshift range $0.2\leq z \leq 1.0$, monitoring the redshift evolution
in \bluefrac in a continuous way.

\begin{figure*}
\centering                  
\includegraphics[width=12cm,angle=270]{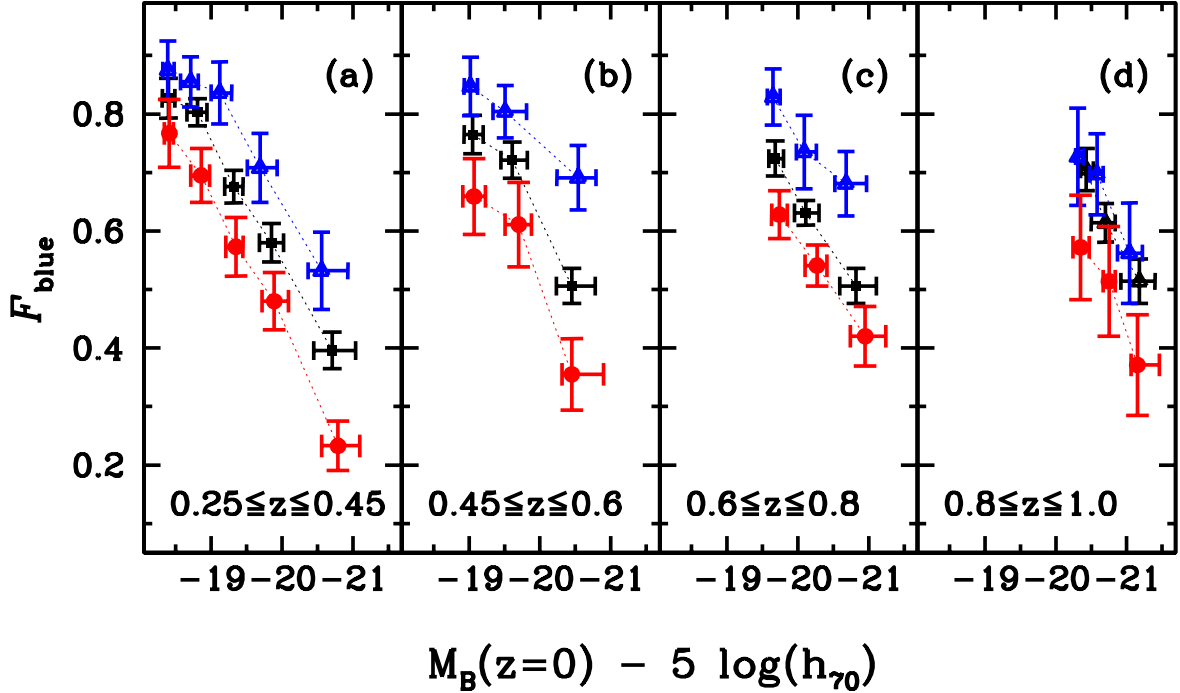}
\caption{The four panels show \bluefrac in different redshift bins, as
indicated on the bottom of each panel, as a function of absolute
luminosity, evolution corrected to $z \sim 0$ to ensure that similar
galaxies are being compared across different redshifts bins. Different
colors refer to different galaxy samples: red circles refer to group
galaxies, blue triangles to isolated galaxies, while black squares to
the total galaxy population. The errors on \bluefrac are obtained
using bootstrapping, while the error bars along the luminosity axis
link the upper and lower quartiles of the luminosity distribution of
galaxies within the bin considered. At all magnitudes and at all
redshifts groups contain less blue galaxies than the field and the
isolated galaxies population. For all redshift bins considered and
irrespective of the environment fainter galaxies are always bluer than
brighter galaxies. For all environments \bluefrac increases with
redshift.}
\label{fig:fbflum}  
\end{figure*}

As a first step, we explored how \bluefrac varies with galaxy
luminosity. We defined four independent redshift bins as shown in
Fig.~\ref{fig:fbflum}: [0.25:0.45], [0.45:0.6], [0.6:0.8],
[0.8:1.0]. Within each of these redshift intervals and using the
volume-limited samples defined in Table~\ref{tab:vollimnumb}, we
defined sub-samples of galaxies in independent bins of galaxy
luminosities. The binning in galaxy luminosity was chosen in such a
way to ensure a sizeable number of galaxies in each environment and
redshift bin considered. \bluefrac and its error bar were estimated
using the procedures described in Section~\ref{sect:sampleanalysis},
while the error bars drawn along the luminosity axis link the upper
and lower quartiles of the luminosity distribution of galaxies within
each bin.

Figure~\ref{fig:fbflum} shows the results obtained for the different
galaxy samples: red circles for group galaxies, blue triangles for
isolated galaxies, and black squares for the total galaxy population.
In each redshift bin all the different galaxy populations display a
decrease in the fraction of blue galaxies for increasing rest-frame
galaxy luminosities, while at a fixed luminosity bin, blue galaxies
are always less common in the group environment than in the field and
most common among the isolated galaxy
population. Figure~\ref{fig:fbflum} therefore suggests that at all
redshifts explored the color of galaxies at a given luminosity becomes
redder earlier in groups than in the field or in lower density
regions.

Furthermore, the differences between the galaxy population of the
three different environments seem to increase at higher luminosities
in each of the four panels of Fig.~\ref{fig:fbflum} and this result
echoes a similar one in \citet{Cucciati2006}.

Towards redshift $z \sim 1$ the differences among the three
environments progressively decrease. However up to the highest
redshift bin explored we do not see any hint of a possible reversal of
the trend of \bluefrac as a function of luminosity, a robust result as
our sample is free from significant color-dependent incompleteness up
to $z\sim1$ (see Section~\ref{sect:VolLim}). Such possible trend
reversal was tentatively detected by \citet{Gerke2007}, albeit with
large error bars, for the redshift bin $0.7 \leq z \leq 1.0$ and for
magnitudes brighter than $M_{B} \sim -21.5$.

\begin{figure*}
\sidecaption                   
\includegraphics[width=12cm,angle=0]{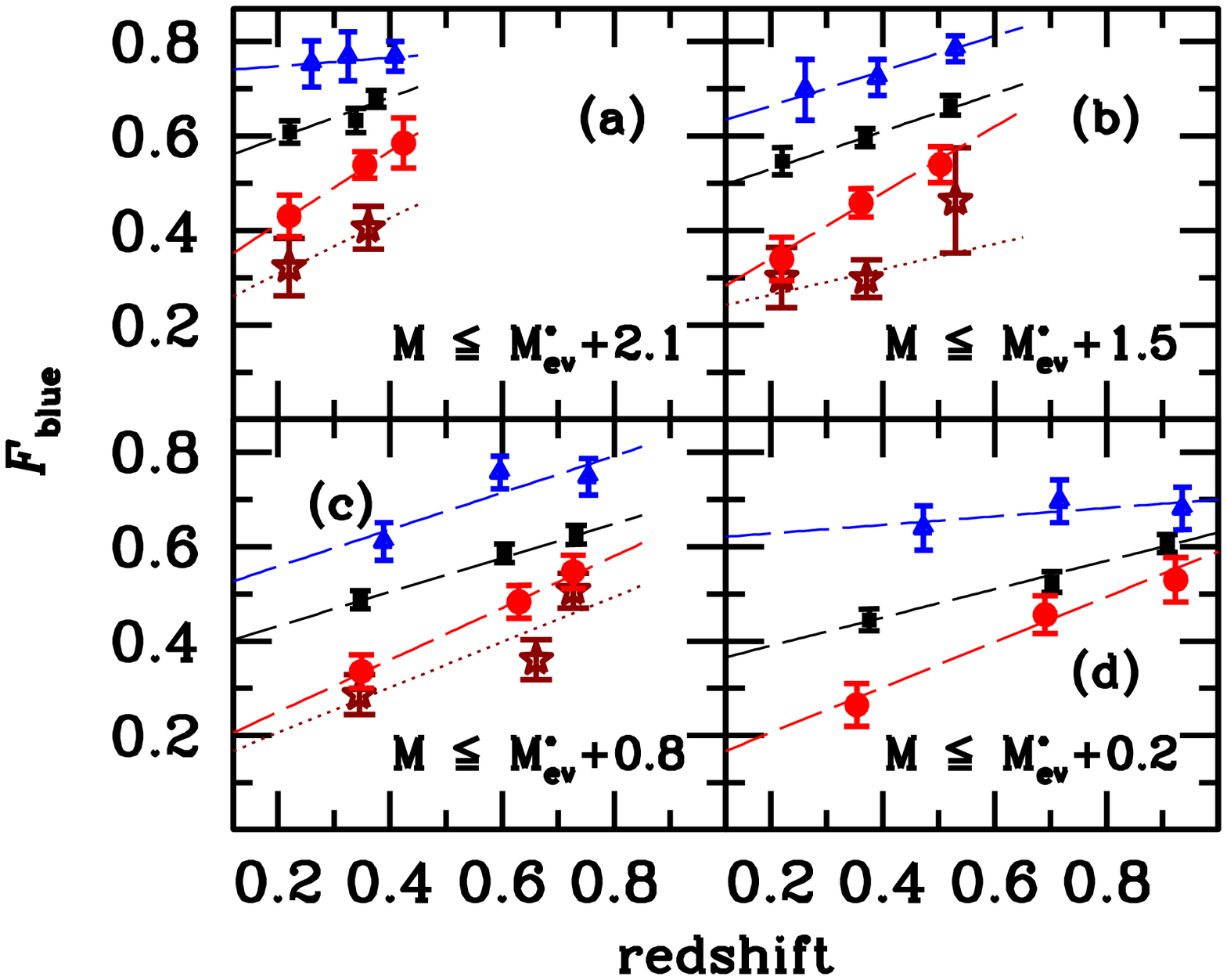} 
\caption{The four panels show \bluefrac as a function of redshift for
each of the different volume-limited samples defined in
Table~\ref{tab:vollimnumb}. The label in each panel indicates the
range in evolution corrected, present-day, absolute magnitude for the
galaxies plotted (we assumed $M^*_{\it ev} = -20.3 - 5~log~h_{70}
-1.1~z$). Red circles refer to group galaxies, blue triangles to
isolated galaxies, while black squares to the total galaxy
population. Brown stars are those corresponding, for each
volume-limited sample, to the population of galaxies in groups with at
least two members in Sample IV, that is sample Group galaxies II in
Table~\ref{tab:vollimnumb}. Color segregation is already in place at
redshift $\sim 1$ and increases sensibly moving from higher to lower
redshifts for all the volume limited samples considered. See text for
more details.}
\label{fig:fbz} 
\end{figure*}

We used the four volume-limited samples and the three galaxy samples
defined in Table~\ref{tab:vollimnumb} to explore in better detail the
redshift trends implied by Fig.~\ref{fig:fbflum}. For each of these
samples, we plotted \bluefrac as a function of redshift in
Fig.~\ref{fig:fbz}, to help determine directly whether the rate of
variation in \bluefrac differs significantly in groups compared to the
field/isolated galaxy population.  Each panel refers to a
volume-limited sample defined by the labels at its bottom, where red
circles indicate \bluefrac for group galaxies, while black squares and
blue triangles show the same quantity for field and isolated galaxies,
respectively.

The first piece of information conveyed by Fig.~\ref{fig:fbz} is that
color segregation appears to be already in place at $z \sim 1$: panel
(d) shows that even in the highest redshift bin explored there is a
small, but significant, difference in \bluefrac among the different
galaxy samples, mirroring the information provided by panel (d) of
Fig.~\ref{fig:fbflum}. Furthermore for each of the luminosity bins
explored color segregation increases with cosmic time, as the
differences of \bluefrac in the group, field and isolated galaxy
populations increase significantly moving from high to low redshifts.

These results are in good agreement with those from the VVDS survey
presented by \citet{Cucciati2009a}.  However we seem to detect
evolution in \bluefrac across the range $0.75 \leq z \leq 1.0$, in
contrast with the results of \citet{Gerke2007} using DEEP2
data-set. We note that comparing directly our panel (d) of
Fig.~\ref{fig:fbz} with the first panel of their Fig. 7, where the
magnitude ranges explored are quite similar and the sample analyzed is
purely volume-limited as in our analysis, the disagreement is not so
evident.

We chose to parametrize the evolution in \bluefrac with redshift with
a law of the form \bluefrac $ \propto (1+z)^{\beta}$. The results of
the best fit solutions obtained with this parametrization are given in
Table ~\ref{tab:vollimfit} and shown as dashed lines in
Fig.~\ref{fig:fbz}.  These lines tend to diverge between high and low
z. Toward higher redshift, one can consider whether, irrespective of
the environment considered, most galaxies in the luminosity ranges
explored resided in the blue cloud, while the red sequence remained
more or less empty. As cosmic time increases, the blue cloud may then
become progressively depleted and the rate at which this depletion
occurs seems to be higher in higher density environments, implying
that the star-formation rate is declining more rapidly in groups and
clusters.

While the extrapolated values of \bluefrac at $z \sim 0$ vary as a
function both of environment and the luminosity cut-off considered,
the values of $\beta$ do not exhibit any appreciable differences
between the different environments as a function of the chosen
luminosity cut-off. On the other hand, there is a noticeable increase
in the value of $\beta$ moving from isolated to group galaxies,
although the error bars are quite large.

Since $\beta$ implies that there is a fractional decrease in \bluefrac
with cosmic time (or, alternatively, a fractional increase of the
percentage of red galaxies with cosmic time) this result suggests that
we detect the signature of an environmental dependence of the
variations in \bluefrac with cosmic time. However the mechanisms
responsible for the environmental trends that we witness cannot be
accurately constrained.

Our results could be the consequence of physical mechanisms operating
in the denser group environment or simply the result of an {\it ab
initio} bias relating galaxy luminosity/mass and its environment. In
other words, we could be witnessing the more rapid quenching of star
formation - and, as a consequence, the faster build-up of the red
sequence - in denser environments, or the delayed and more efficient
replenishing of the blue cloud in lower density environments.  We will
return to this point in the following sections. 

\begin{table*}
\caption{Summary of fits results for \bluefrac as a function of
redshift in the volume-limited samples defined in
Table~\ref{tab:vollimnumb}. We parametrized the evolution of \bluefrac
with a fit of the form \bluefrac $ \propto (1+z)^{\beta}$.
We assume $M^*_{\it Ev} = -20.3 - 5\times log~(h_{70}) -1.1\times z$.} 
\label{tab:vollimfit}  
\centering                  
\begin{tabular}{c c c c c c c c c }    
\hline\hline                 
                & \multicolumn{2}{c}{Sample I}  & \multicolumn{2}{c}{Sample II} & \multicolumn{2}{c}{Sample III} & \multicolumn{2}{c}{Sample IV}  \\   
 $M_{B}$ range  & \multicolumn{2}{c}{$M_{B}\leq M_{B \it Ev}^*+2.1$} & \multicolumn{2}{c}{$M_{B}\leq M_{B \it Ev}^*+1.5$} & \multicolumn{2}{c}{$M_{B}\leq M_{B \it Ev}^*+0.8$} & \multicolumn{2}{c}{$M_{B} \leq M_{B \it Ev}^*+0.2$} \\
    $z$ range   & \multicolumn{2}{c}{$0.1\leq z\leq0.45$} & \multicolumn{2}{c}{$0.1\leq z\leq0.6$} & \multicolumn{2}{c}{$0.1\leq z\leq0.8$} & \multicolumn{2}{c}{$0.1\leq z\leq 1.0$} \\                  
\hline  \hline                    
                   & \bluefrac($z=0)$  & $\beta$ & \bluefrac($z=0)$ & $\beta$ & \bluefrac($z=0)$ & $\beta$ & \bluefrac($z=0)$ & $\beta$ \\
All galaxies       & 0.51$\pm$0.06 & 0.88$\pm$0.40 & 0.45$\pm$0.04 & 0.92$\pm$0.27 & 0.36$\pm$0.03 & 0.99$\pm$0.20 & 0.32$\pm$0.03  & 0.96$\pm$0.18  \\
Isolated galaxies      & 0.73$\pm$0.15 & 0.16$\pm$0.67 & 0.58$\pm$0.10 & 0.69$\pm$0.44 & 0.48$\pm$0.08 & 0.85$\pm$0.36 & 0.60$\pm$0.12  & 0.23$\pm$0.36  \\ 
Group galaxies I   & 0.29$\pm$0.08 & 2.01$\pm$0.87 & 0.24$\pm$0.06 & 2.05$\pm$0.67 & 0.19$\pm$0.05 & 1.97$\pm$0.50 & 0.16$\pm$0.05  & 1.87$\pm$0.53  \\ 
Group galaxies II  & 0.21$\pm$0.12 & 2.09$\pm$1.85 & 0.19$\pm$0.08 & 1.75$\pm$1.27 & 0.14$\pm$0.05 & 2.20$\pm$0.61 & 0.16$\pm$0.05  & 1.87$\pm$0.53  \\ 
\hline                             
\end{tabular}
\end{table*}

The group points in Fig.~\ref{fig:fbz} do not correspond to a group
population that is homogeneous across each of the four different
panels but to the samples indicated as Group galaxies I in
Table~\ref{tab:vollimnumb}, \ie galaxies in groups of more than two
members observed within the volume-limited sample plotted in each
panel. Moving from panel (a) to panel (d) in Fig.~\ref{fig:fbz}, we
consider groups that are intrinsically more rich, since they possess
two or more members at progressively brighter cut-off magnitudes.
Accordingly the observed decrease in \bluefrac for the group galaxy
population between the first and the last panel of Fig.~\ref{fig:fbz}
is the result of two different effects: the brightening of the galaxy
population, an effect easily visible also for the isolated and the
field samples, and the increasing richness of groups observed in the
brighter volume-limited samples.

It is therefore interesting to remove the richness-dependent effect
from this plot and to compare galaxies residing in groups of
homogeneous richness across the different volume-limited samples,
using the sample of groups with two or more members brighter than
$M_{B} \leq M_{B \it Ev}^*+0.2$, \ie the brightest absolute magnitude
cut-off in our sample.  This way, we select for each panel, except
panel (d) whose group sample remains unchanged by definition, a set of
groups richer than those considered before. Obviously, once a group
survives in the catalogue defined by the more stringent luminosity
cut-off, we then plot in the appropriate panel all its members
observed within the volume-limited sample under study.  The result of
this exercise is shown by the brown stars in Fig.~\ref{fig:fbz} and
by the last row of Table~\ref{tab:vollimfit}, labeled Group galaxies
II.  In this case, the groups considered are homogeneous in richness
up to $z \sim 1$.  The difference compared to the field population
(and the isolated galaxies) increases significantly for these richer
groups. In contrast the dependence of \bluefrac on the rest-frame B
magnitude of the group population is reduced significantly.
Table~\ref{tab:vollimfit} shows that the slopes of the fits to the
Group galaxies II points are virtually indistinguishable among
themselves, irrespective of the luminosity limits adopted in each of
the four different panels.

\section{Blue fraction as a function of group properties to $z \sim 1$}  
\label{sect:GroupsSamples} 

The results obtained in the previous section using Group galaxies II
samples suggest that group richness is an important ingredient in
setting the value of $\it{F_{blue}}$, possibly more influential than
the galaxy rest-frame B magnitude.

Group richness can be considered, albeit with a large scatter, a proxy
for the mass of the halo where the group resides (see Knobel et al.,
2009). Therefore, the results just obtained echo at higher redshifts
findings in the local Universe by \citet{Weinmann06}. Using both
galaxy colors and specific star formation rate indicators to define
samples of early/late type galaxies, these authors showed that at
fixed halo mass the dependence of the galaxy type fraction on
luminosity is quite weak. 

Concerning the dependence of \bluefrac on group richness and/or
velocity dispersion - another possible proxy of halo mass - many
conflicting observational results, however, exist in the
literature. Some authors have claimed the presence of a relationship
between group and galaxy properties \citep[see, \eg][to quote a
few]{Biviano1997, ZabludoffMulchaey1998a, Margoniner2001,
Martinez2002a, Goto2003, Tanaka2004, Poggianti2006, DeLucia2007,
Gerke2007, Koyama2007}, while other authors have claimed that such
relationships are not present \citep[see, \eg][]{Smail1998,
Ellingson2001, Fairley2002, DePropris2004, Goto2005c, Wilman2005a,
Popesso2007}.

In the following, we explore with our sample the dependence of
\bluefrac on group richness and velocity dispersion.

\subsection{Blue fraction as a function of group richness}  
\label{sect:GroupRich}   

The use of the term ``richness'' for galaxy clusters dates back to
Abell, who introduced a broad classification of clusters in three
richness classes based on counting galaxies between m3 and m3 + 2 mag,
where m3 is the magnitude of the third-brightest galaxy
\citep[see][]{Abell1958}. In this paper, we use the term richness for
each group to simply indicate the number of members observed in each
of the different volume-limited samples. As such, the richness of a
group is not an absolute number, but varies depending on the absolute
magnitude cut-off chosen to define the sample of groups.  A possible
better name for this quantity could be group {\it numerosity}. In
estimating group richness/numerosity, however, even within this more
limited definition, one has to consider corrections to the number of
galaxy members observed, to properly account for the large-scale
variations in the mean sampling rate of the \10K.

While the mean sampling rate of the \10K is about $\sim 30$\% and
increases up to $\sim 40$\% in the restricted central area adopted for
our analysis, there are large spatial variations in this number (see
Fig.~\ref{fig:radec_distr_map}).

To correct for this problem, we estimated richness using the weighting
scheme discussed in Section~\ref{sect:IncomplCorr}. For each volume
limited sample we simply added galaxy weights to count the group
members, by writing richness as ${\cal N}_{corr} = \sum_{j=1}^M w_j$,
where M is the number of members observed in each volume-limited
sample.

\begin{figure}
\includegraphics[width=9cm,angle=0]{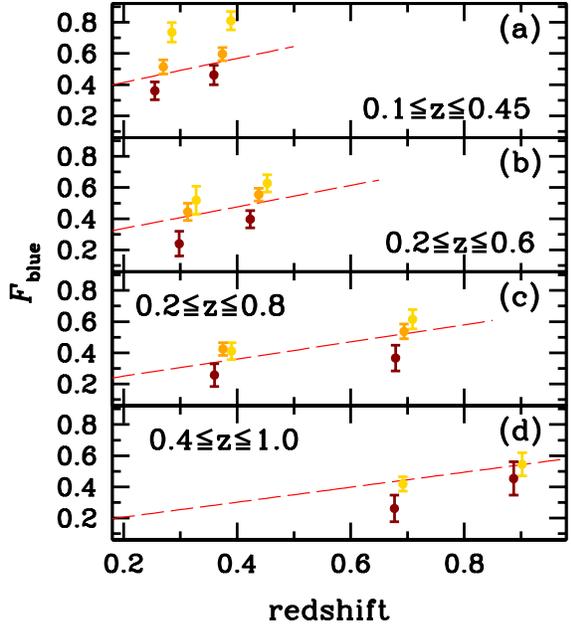} 

\caption{Dependence of the blue fraction \bluefrac on group redshift
and richness.  In each panel, the different redshift ranges indicated
by the label are considered. In panels (a), (b) and (c), we plot the
value of \bluefrac for groups with $N_{corr} \leq 4$, yellow points,
$4 < N_{corr} \leq 10$, orange points and $N_{corr} > 10$, brown
points, in samples I, II and III respectively.  In panel (d) we plot
the value of \bluefrac for groups in sample IV with $N_{corr} \leq 8$,
yellow points and $N_{corr} > 8$ brown points. The points have been
slightly offset in redshift for the sake of clarity. In all panels,
the red dashed line corresponds to fits obtained for the entire galaxy
group population, as shown in Fig.~\ref{fig:fbz}.  There is a
consistent trend for all the samples and all the redshift bin
explored: groups with higher $N_{corr}$ tend to have a lower fraction
of blue galaxies, a trend superimposed on the global decrease of
\bluefrac moving from high to low redshift.}
\label{fig:rich_frac}
\end{figure}

Figure~\ref{fig:rich_frac} shows the dependence of the value of
\bluefrac on group richness. In each panel, we consider different
redshift ranges, corresponding to the four volume-limited samples
defined in Table~\ref{tab:vollimnumb}, and divide the corresponding
group sample according to richness. In all panels, the red dashed line
indicates the fit to the global galaxy population obtained in
Section~\ref{sect:VolLimSamples}.  In the first three panels, the
yellow points correspond to groups of observed richness ${\cal
N}_{corr} \leq 4$, the orange points to groups of observed richness $
4 < {\cal N}_{corr} \leq 10$ and the brown points to groups of
observed richness $ {\cal N}_{corr} > 10$. In panel (d), yellow points
correspond to groups of observed richness ${\cal N}_{corr} \leq 8$ and
brown points to groups of observed richness ${\cal N}_{corr} > 8$. The
limits chosen to divide the groups into bins of richness are arbitrary
but ensure that each richness bin contains a sizeable number of group
member galaxies (always above $\sim 30$).
 
The main result presented in Fig.~\ref{fig:rich_frac} is that for all
redshifts explored and volume limited samples, \bluefrac decreases
monotonically between higher and lower redshifts and richer groups
have a lower value of $\it{F_{blue}}$. This result is in agreement with
similar trends observed in the local Universe,
\citep[see][]{Margoniner2001, Goto2003}.

We noted in Section~\ref{sect:sample10Kgroups} that in our sample we
do not have any more massive relaxed clusters, especially at low
redshift, but instead we probe mainly the poorer clusters and group
environment. For our lower richness groups, the results obtained --
especially in the lowest redshift bin, where we are dealing with the
poorest groups of the sample in absolute terms -- blend with those
observed for the global field population. We may be concerned that
these results are just the by--product of a higher interloper fraction
for these (poorer) groups. However, as discussed in detail in Knobel
et al. (2009), the interloper fraction of our group catalogue shows a
minimal increase when the number of detected members
decreases. Furthermore in the local Universe extremely poor groups are
known to be dominated by spiral galaxies
\citep{ZabludoffMulchaey1998a}, and so what we are witnessing is most
probably a real physical trend.

\subsection{Blue fraction as a function of group velocity dispersion} 
\label{sect:GroupVeldisp}   

We binned our group sample using also the observed line-of-sight
velocity dispersion $\sigma$, another possible proxy for group mass.
The estimate of $\sigma$ is difficult, especially when one is dealing,
as in our case, with groups of only a few members with measured
redshifts \citep[see][]{Beers1990}. However, restricting the analysis
to groups with observed numbers of members equal or greater than 5
allows us to observe a reasonable correlation between $\sigma$ and
halo group mass (Knobel et al., 2009).  We therefore used only groups
with at least 5 members with measured redshifts in the \10K to explore
how \bluefrac depends on the measured $\sigma$ in different redshift
bins and for different volume-limited samples.

\begin{figure}
\includegraphics[width=9cm,angle=0]{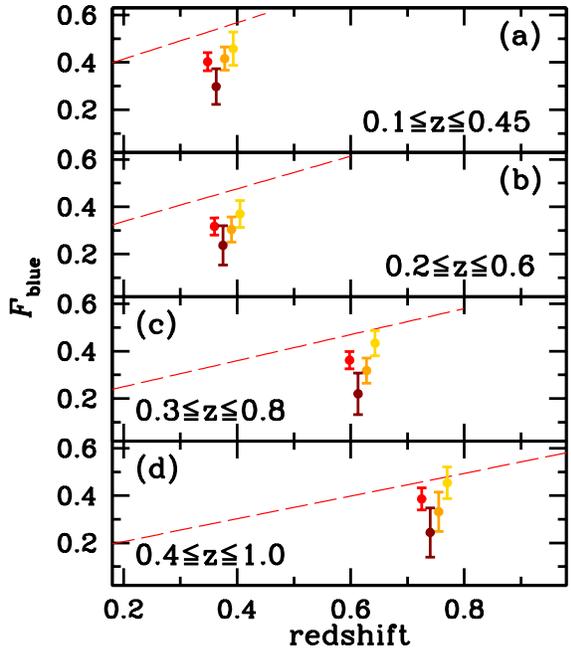} 
\caption{Dependence of the blue fraction \bluefrac on redshift and
line-of-sight velocity dispersion of groups.  In each panel, the
different redshift ranges indicated by the label are considered. In
all panels only groups detected with at least 5 members in the
flux-limited sample used by the group detection algorithm are plotted,
to avoid the large uncertainty in velocity dispersion measurement when
poorer structures are considered. In panel (a), the yellow point shows
\bluefrac for groups with $\sigma \leq 250$ \kmsec, the orange point
for groups with $250 \leq \sigma \leq 550$ \kmsec, and the brown point
for groups with $\sigma > 550$ \kmsec.  In the three remaining panels,
the yellow points show \bluefrac for groups with $\sigma \leq 350$
\kmsec, the orange points for groups with $350 \leq \sigma \leq 650$
\kmsec, and the browns point for groups with $\sigma > 650$ \kmsec. In
all panels, the red dashed line corresponds to fits obtained for the
entire galaxy group population in the volume-limited sample, as in
Fig.~\ref{fig:rich_frac}, while the red points are those corresponding
to the sample of groups detected with at least 5 members in the
flux-limited sample, irrespective of their measured line-of-sight
velocity dispersion. In each plot, the points are located in redshift
at the median value of the sample considered, with a small offset, for
the sake of clarity, among the different $\sigma$ limited
samples. There is a consistent trend for all the redshift bins
explored: groups with higher line-of-sight velocity dispersion tend to
have a lower fraction of blue galaxies. }
\label{fig:slos_frac} 
\end{figure}

Our results are shown in Fig.~\ref{fig:slos_frac}. In panel (a), the
yellow point indicates \bluefrac for groups with $\sigma \leq 250$
\kmsec, the orange point for groups with $250 \leq \sigma \leq 550$
\kmsec, and the brown point for groups with $\sigma > 550$ \kmsec.  In
the three remaining panels, the yellow points show \bluefrac for
groups with $\sigma \leq 350$ \kmsec, the orange points for groups
with $350 \leq \sigma \leq 650$ \kmsec, and the brown points for
groups with $\sigma > 650$ \kmsec. In all panels, the red dashed line
corresponds to the fits obtained for the entire galaxy groups
population in the volume-limited sample, as in
Fig.~\ref{fig:rich_frac}, while the red points are those corresponding
to the sample of groups detected with at least 5 members in the
flux-limited sample irrespective of their velocity
dispersion. Figure~\ref{fig:slos_frac} shows that there is a
consistent trend in all panels: groups with higher velocity dispersion
tend to have a lower fraction of blue galaxies.

Limiting the sample in the analysis only to groups of at least 5
observed members, produces the systematic offset towards lower \bluefrac
observed in each of the panels when comparing the red points to the
dashed red lines: with the requirement of $N \geq 5$ we remove
from the sample the poorer groups in each volume-limited sample.

We conclude this section with a word of caution. One would be
tempted to compare points as a function of redshift across the last
three different panels, since the $\sigma$ cut-off chosen is equal for
each of them, and after showing in Section~\ref{sect:isolgrfield} that
galaxy luminosity is of much less importance in determining \bluefrac
than group richness.  This comparison would apparently result in a
statement of {\it non-evolution} of \bluefrac as a function of
redshift for groups of similar velocity dispersion.

However one needs to be extremely careful when comparing results
obtained at different redshifts. As cosmic time increases a system
will experience an increase in its velocity dispersion, as its halo
mass becomes higher due to structure growth (see, \eg the
prescriptions obtained by using semi-analytic models by
\citet{Wechsler2002} and \citet{Poggianti2006} for an application).
Furthermore, one should keep in mind that the imposed cut-off on the
number of members observed in the flux-limited sample is expected to
introduce a strong bias favoring richer groups moving from panel (a)
to panel (d), as the absolute luminosity of the observed galaxies
becomes brighter. What appears as an absence in evolution of \bluefrac
as a function of redshift at a fixed velocity dispersion is thus at
least partly caused by the progressive bias against lower richness
groups moving from panel (a) to panel (d).

Estimates of group richness and group velocity dispersion often have
large error bars, due to the paucity of member-galaxy samples
available, and group richness/group velocity dispersion are properties
that have a large scatter in their relationship to more fundamental
quantities - as the mass of the halo where the group resides. As a
consequence, it is unsurprising to observe a large scatter in the
trends that relate group richness and group velocity dispersion with
the value of $\it{F_{blue}}$.

To avoid producing biased results on evolution proper care has to be
taken to compare properties of group samples that are truly
homogeneous in the different redshift bins explored.

\section{Moving from luminosity to stellar mass: redefining the Butcher-Oemler effect} 
\label{sect:RedefBO}   

The results obtained in the previous section can all be interpreted in
the framework of the classical Butcher-Oemler effect
\citep{ButcherOemler1978}, in its wider context extended to group
population \citep[see][]{Allington-Smith1993}.

Taking advantage of the wide redshift and galaxy/group population
coverage of the \10K galaxy/groups catalogue, we have been able to
show that the blueing of the galaxy population in groups/clusters when
moving to higher redshift, originally observed by these authors thirty
years ago, is a real effect, which differs from that observed for the
global galaxy population. It also exhibits specific trends as a
function of both galaxy B-band rest-frame luminosity and group
properties. These trends are present as a function of environment in
all luminosity and redshift bins explored and seem to become
progressively more conspicuous moving from $z \sim 1 $ to lower
redshifts and from lower to higher luminosities. 

However almost galaxy properties depend strongly on galaxy stellar
mass, and this is true particularly for galaxy colors. Galaxy stellar
mass in turn is known to correlate with environment and can be a key
player in determining galaxy properties and in linking them to the
environment in which they reside.

It is therefore important to check whether the strong effects evident
in luminosity-selected samples are still present when the analysis is
repeated using stellar-mass-selected samples. In this way, we probe
the possibility of these effects being the distorted/amplified
reflection -- related to the biased view imposed by the luminosity
selection -- of more fundamental relationships either between masses
and environment, or between masses and galaxy colors.

In the following Sections, we re-examine the original Butcher-Oemler
results using mass-limited samples instead of volume-limited
samples. What becomes of the observed strong trends in \bluefrac as a
function of environment and redshift, shown in, \eg Fig.~\ref{fig:fbz},
when one utilizes samples complete in mass? Are we able to confirm the
existence of a higher proportion of blue galaxies in higher redshift
groups with respect to their lower redshift counterparts, when using
mass-limited samples? Can we still see an excess of red galaxies in
groups with respect to the field/isolated galaxy population even when
analyzing mass bins?

Obviously analyzing volume-limited stellar-mass selected samples
involves a significant reduction in the galaxy sample size available
to our study, as the selection of mass-complete samples implies the
rejection of a large number of low-mass galaxies for which our
$I_{AB}$ selected redshift survey is incomplete. However this is an
unavoidable step in clarifying the key mechanisms determining the
relationships observed for luminosity-selected samples.
 
Galaxy stellar mass has the further advantage of being more {\it
stable} than its luminosity. The B-band galaxy rest-frame luminosity
may indeed change dramatically during a galaxy lifetime because of
bursts of star-formation. Even in the absence of these bursts, the
rest-frame B-band luminosity evolves with redshift, possibly in
different ways for different galaxy populations, and one needs to
introduce - as we have - an average evolution correction term to
sample homogeneous galaxy populations in the different redshift bins
explored. On the other hand, stellar mass varies to a far lesser
extent during a galaxy's life; it also increases due to star formation
and mergers, but by a smaller percentage, as confirmed both by
observational evidence, showing that up to $z\sim1$ the mass function
evolves only mildly (see \citet {Pozzetti2007} and references
therein), and by numerical simulations, \citep{DeLucia2006} . As a
consequence, the selection of mass-limited samples eases the task of
tracing the same population of galaxies in the different redshift bins
explored.  In the subsequent Sections, we will investigate the impact
of the use of mass-selected samples on our analysis.

\section{Defining stellar-mass, volume-limited samples} 
\label{sect:masscompl}   

To construct volume-limited, stellar-mass selected samples, we
followed a simple approach. For each of the four redshift bins adopted
in the previous analysis, we estimated the limiting mass at which even
the oldest/reddest galaxies (\ie those with the maximum possible
stellar mass-to-light ratio) would be observable given the magnitude
limit of our survey.

To estimate this limiting mass, we first proceeded by calculating the
limiting stellar mass of each galaxy, \ie the stellar mass it would
have, at its spectroscopic redshift, if its apparent magnitude were
equal to the limiting magnitude of our survey ($I_{AB} = 22.5$).  We
then used these estimated limiting masses to define, in bins of
$(U-B)$ rest-frame colors for each redshift bin, the mass ${\cal
M}_{cut-off}$ below which $85$\% of galaxies of that color lie.  The
value of ${\cal M}_{cut-off}$ for the reddest galaxies in each
redshift bin is the one that we use as limiting mass.

\begin{figure}
\includegraphics[width=9cm,angle=0]{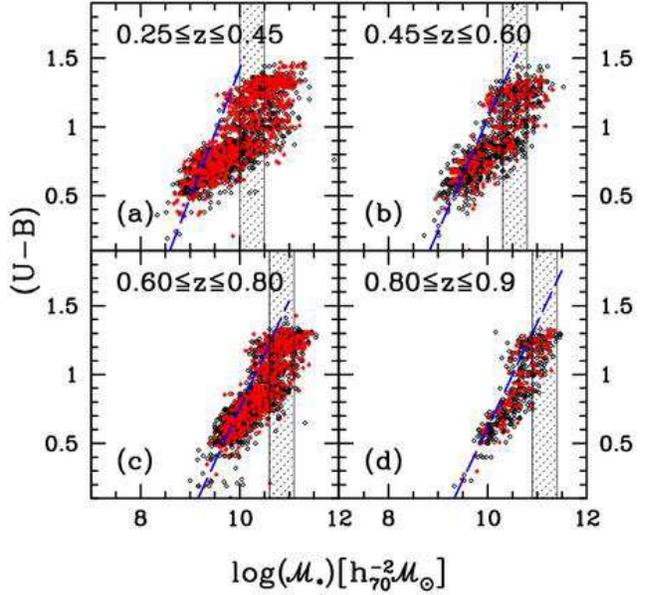} 
\caption{ Rest-frame (U-B) colors plotted versus galaxy stellar mass
in solar mass units for the samples defined in Table
\ref{tab:vollimnumb}.  On top of each panel, we write the redshift
bins considered. For each panel, only galaxies contained in the
corresponding volume-limited sample have been plotted (sample I to
sample IV). The points in red show galaxies located in groups, while
those in black correspond to the total population of galaxies. The
blue dashed line in each panel indicates the color-dependent $85$\%
completeness mass-limit (see text for more details on how it is
computed). The shaded area indicates the mass bins that we used in our
analysis, and its lower boundary in mass equals the limiting mass
above which all galaxies, irrespective of their color, are observed in
our flux-limited survey.}
\label{fig:masslim}
\end{figure}

Figure~\ref{fig:masslim} shows the distribution of $(U-B)$ colors
versus stellar masses for each of the volume-limited samples defined
in Table~\ref{tab:vollimnumb}. The points in red show galaxies located
in groups, while those in black correspond to the total population of
galaxies. The blue, dashed line in each panel indicates the fit to the
values of the color-dependent $85$\% completeness mass-limit,
estimated as described above. The lower boundary to the shaded
rectangular area in each panel indicates the limiting mass above which
all galaxies, even those redder in colors, are observed in our
magnitude-limited survey. Figure~\ref{fig:masslim} confirms that
adopting mass volume-limited samples rejects a large number of
lower-mass, bright, blue galaxies, which were included in B-band
luminosity, volume-limited samples. 

For each of the mass volume-limited samples
Table~\ref{tab:masslimnumb} summarizes its lower mass limit and the
number of galaxies contained in both the full galaxy sample and
isolated/group galaxy samples.

\begin{table*}
\caption{Summary of the four mass volume-limited data
samples. {\it Mass} is in units of $log({\cal M}_*/({\cal M}_{\odot} \times h_{70}^{-2}))$. }
\label{tab:masslimnumb}  
\centering                  
\begin{tabular}{c c c c c }    
\hline\hline                 
                & Sample M-I  & Sample M-II & Sample M-III & Sample M-IV  \\   
Stellar mass range  & \it{Mass}$\geq 10.0 $ & \it{Mass}$\geq 10.3 $ & \it{Mass}$\geq 10.6 $ & \it{Mass}$\geq 10.9 $ \\
    $z$ range   & $0.1\leq z\leq0.45$ & $0.1\leq z\leq0.6$ & $0.1\leq z\leq0.8$ & $0.1\leq z\leq 0.9$ \\                  
\hline  \hline                    
All galaxies       &     883    &   914     &  1033    &   491  \\      
Isolated galaxies  &     119    &   141     &   131    &    55  \\
Group galaxies I   &     386    &   355     &   330    &   137  \\
Group galaxies II  &     155    &   165     &   230    &   137  \\
\hline                             
\end{tabular}
\end{table*}

\begin{figure*}
\centering                  
\includegraphics[width=10cm,angle=270]{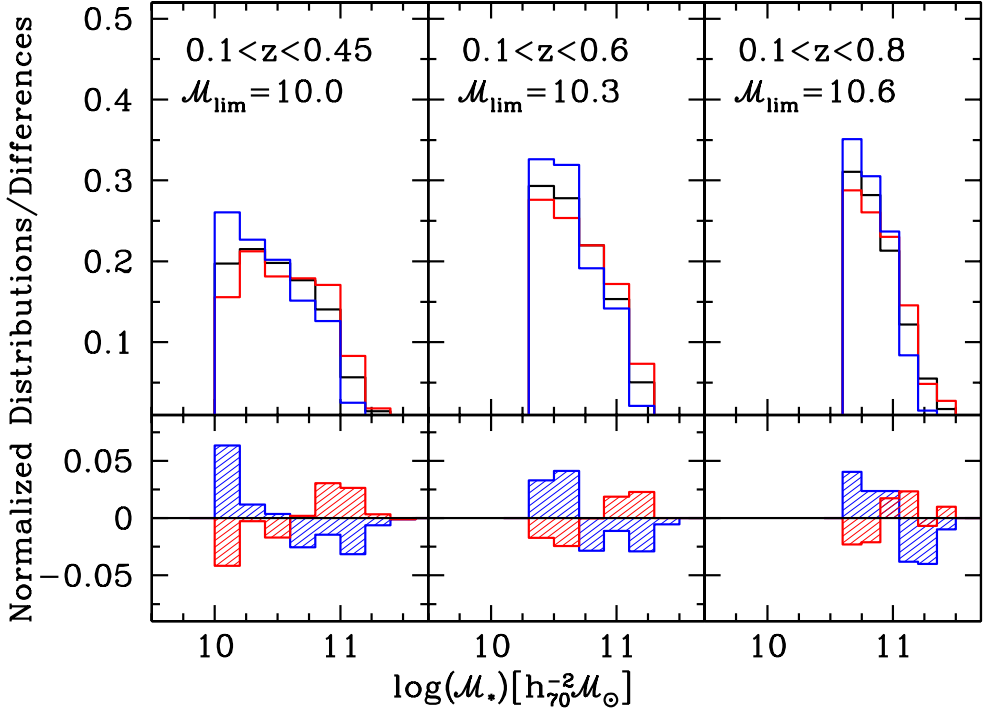} 
\caption{Top panels: Normalized histograms of galaxy mass for
different galaxy populations.  Isolated galaxies are in blue, group
galaxies are in red and black galaxies are the full sample. Labels on
top indicate the redshift bin considered and the cut-off mass
adopted. Bottom panels: Histogram of normalized differences with
respect to the full galaxy sample of the group population (shaded in
red) and the isolated galaxy population (shaded in blue).  With
respect to the whole galaxy population there is a visible and
statistically significant excess of low/high mass galaxies in the
isolated/group galaxy sample.}
\label{fig:histo_masses} 
\end{figure*}

Comparing the numbers in Table~\ref{tab:masslimnumb} with those in
Table~\ref{tab:vollimnumb}, it is clear that the samples that become
more depleted moving from a luminosity to a mass selection are those
of isolated galaxies. On average, these samples experience a decrease
in number by a factor of $\sim 3 $. The group samples, instead, are at
most just halved. The blue low-mass galaxies that are excluded when
moving from luminosity to mass limited samples are a larger fractions
of galaxies residing in low-density environments. It is natural to
conclude that at least part of the strong trends of \bluefrac as a
function of environment observed in our volume-limited samples are
driven by the large population of lower mass, bright blue galaxies for
which we miss the redder, equally low mass, counterparts \citep[see]
[for a similar suggestion]{DePropris2003}. In other words, the trends
that we witness in Fig. ~\ref{fig:fbz} are at least partly due to the
bias in B-band magnitudes volume-limited samples against red, low-mass
galaxies, which are too faint to be included by the adopted luminosity
cut-off.

It remains to be seen whether these trends are still observed when
adopting mass volume-limited samples for the analysis, or whether mass
is all that is needed for predicting galaxy colors, irrespective of
environment, a possibility still compatible with our results until
now. Such a possibility is the one expected in simple, pure {\it
nature}, galaxy formation models, where the characteristics of a
galaxy (\eg colors, spin) are primarily determined by the mass of the
dark matter halo in which it resides, which is in turn closely related
to the galaxy stellar mass on one side, and to the density field on a
$\sim 1~Mpc$ scale on the other side
\citep[see, \eg][]{CooraySheth2002}.  In this framework, the color
segregation just mirrors the change in the distribution of galaxy
stellar masses as a function of environment.

\subsection{Mass segregation in groups up to $z \sim 1$}     

The variation in the galaxy stellar mass function between different
environments has been observed in the local Universe \citep[see, \eg]
[]{Baldry2006}, and, at higher redshifts, in DEEP2 data
\citep{Bundy2006}, in VVDS data (Scodeggio et al., 2009), in COSMOS
data \citep{Scoville2007}, and in zCOSMOS \10K data (Bolzonella et
al. 2009). In this section, we check whether mass segregation is
detectable using our group, field, and isolated galaxy samples.

\begin{table*}
\caption{Summary of the four mass volume-limited data samples. 
{\it Mass} is in units of $log({\cal M}_*/({\cal M}_{\odot}\times h_{70}^{-2}))$.} 
\label{tab:massbinnumb}  
\centering                  
\begin{tabular}{c c c c c }    
\hline\hline                 
                & Sample MM-I  & Sample MM-II & Sample MM-III & Sample MM-IV  \\   
 Stellar mass range  & $10.0 \leq {\it Mass}\leq 10.5 $ & $10.3 \leq {\it Mass}\leq 10.8 $ & $10.6 \leq {\it Mass}\leq 11.1 $  & $ 10.9 \leq {\it Mass}\leq 11.4 $ \\
    $z$ range   & $0.1\leq z\leq0.45$ & $0.1\leq z\leq0.6$ & $0.1\leq z\leq0.8$ & $0.1\leq z\leq 0.9$ \\                  
\hline  \hline                    
All galaxies       &     437   &   617     &   885    &   477  \\      
Isolated galaxies  &      64   &   101     &   117    &    45  \\
Group galaxies I   &     174   &   221     &   330    &   132  \\
Group galaxies II  &      56   &    95     &   187    &   132  \\
\hline                             
\end{tabular}
\end{table*}

Figure \ref{fig:histo_masses} shows in its top panels the normalized
histograms of the mass distribution for the first three mass
volume--limited samples of Table~\ref{tab:masslimnumb}, plotted in
red, blue and black for the group, isolated, and all galaxy samples,
respectively. The bottom panels show, shaded in red, the difference
between the group and all-galaxy normalized histograms, and shaded in
blue the difference between the isolated and all-galaxy normalized
histograms. There is a visible excess of both low-mass galaxies in the
isolated galaxy sample and high-mass galaxies in the group galaxy
sample, and the significance of this trend, estimated using a K-S
test, is always at least $\sim 2.3 \sigma$ or more for all
mass/redshift ranges considered. For sample M-IV of
Table~\ref{tab:masslimnumb}, there is no significant difference
between the mass distributions of isolated and group galaxies, a
result possibly caused by both the lower number statistic and the
narrower mass range considered. We therefore have not plotted the
corresponding histograms.

Given these differences in the mass distribution in different
environments, we need to define mass bins that are narrow enough for
mass segregation to become negligible before we can disentangle the
mass/environment influence in determining galaxy colors.  Only in this
way shall we be able to check if environment has truly some influence
on galaxy colors other than being the by--product of mass segregation.
This is the approach we adopt in the following section, using as mass
bins those indicated by the gray shaded areas in
Fig.~\ref{fig:masslim}. A K-S test applied to the mass distribution
within these bins confirms that there is no residual significant
difference in mass among galaxies located in different environments.

\subsection{Blue fraction as a function of galaxy mass and environment up to $z \sim 1$}  
\label{sect:isolgrfield}   

\begin{figure*}
\sidecaption                   
\includegraphics[width=12cm,angle=0]{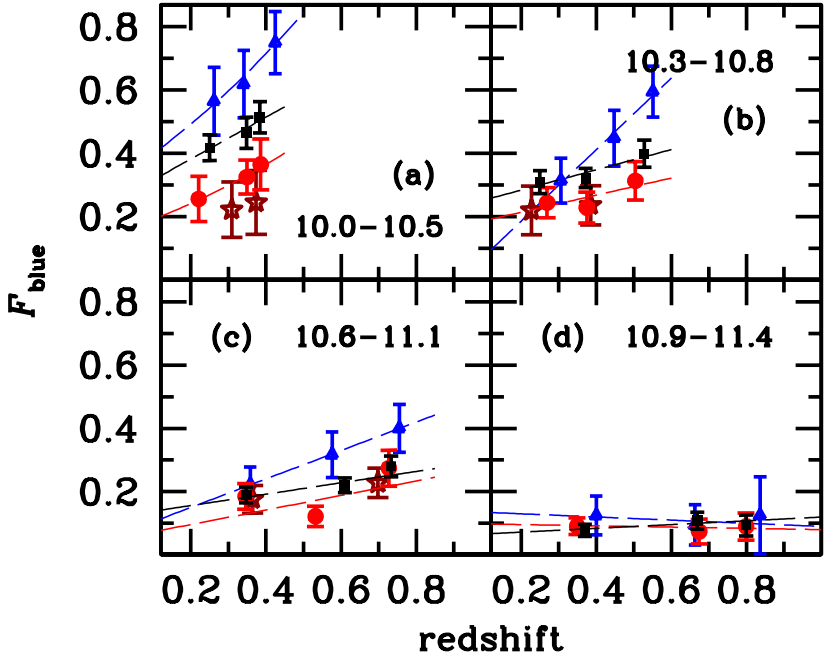}
\caption{ The four panels show \bluefrac as a function of redshift for
each of the different mass limited samples defined in Table
\ref{tab:masslimnumb}, as indicated by the labels.  Red circles refer
to group galaxies, blue triangles to isolated galaxies, and black squares 
to the total galaxy population. Brown stars are those corresponding,
for each of the mass limited sample considered, to the population of
galaxies in groups with at least two members in sample IV. While for
the lowest mass bin explored there is still a significant residual
difference in color as a function of environment, such difference
progressively disappears moving to higher masses.  }
\label{fig:fbz_mass} 
\end{figure*}

We explore how \bluefrac changes as a function of environment in bins
of mass volume-limited samples.  For our analysis, we use the
logarithmic mass bins shown by the shaded rectangles in
Fig.~\ref{fig:masslim}, whose number of galaxies are listed in
Table~\ref{tab:massbinnumb}. Because of the large reduction in our
sample when adopting mass volume-limited samples, we used bins
partially overlapping in mass, a choice dictated by the desire to have
a sufficient number of galaxies in each bin such that our findings
could be deemed statistically significant. As a consequence, the
results shown for the various mass bins are not completely
independent.  Needless to say, the completion of zCOSMOS bright will
enable a much more detailed analysis.

Figure~\ref{fig:fbz_mass} shows the fraction of blue galaxies as a
function of redshift in the four samples and in the three galaxy
samples listed in Table~\ref{tab:massbinnumb}. In each of the four
panels, the red circles show \bluefrac for group galaxies, while the
black squares and the blue triangles show the same quantity for field
and isolated galaxies, respectively. The labels in each panel indicate
the mass range under inspection.

The first information that this plot conveys is that color segregation
is still present in the lowest mass bin explored, the one shown in
panel (a), while in the intermediate-mass bins, \ie in panels (b) and
(c), is barely detectable and only at the highest limits of the
redshift ranges explored. For the highest mass bin, in panel (d),
there is no hint of a color segregation up to $z \sim 1$ and no
evolution with redshift is detectable.

We parametrized the evolution in \bluefrac with a fit of the form
\bluefrac $ \propto (1+z)^{\beta}$, and the results are indicated by
the dashed lines in Fig.~\ref{fig:fbz_mass}. When dealing with samples
defined by bins in mass, however, at odds with what we discussed in
Section~\ref{sect:isolgrfield}, the dashed lines obtained from the fit
to the data points seem, if anything, to indicate that the relative
differences of \bluefrac between group, field, and isolated galaxy
samples progressively disappear moving from high to low redshift.

One can imagine a time when, irrespective of the environment
considered, most galaxies in each of the mass bin ranges explored,
reside on the red sequence, having exhausted their fuel for star
formation, while the blue cloud becomes more or less empty. This seems
to be already the case for the highest mass bin in our plot. Panel (d)
indicates that the majority of red-sequence galaxies in the mass range
$10.9\leq log({\cal M}_*/{\cal M}_{\odot})\leq 11.4$, were already in place, irrespective of
the environment, at the highest redshift bin we can explore ($ z \sim
0.9$).  

In contrast, for the lower mass bins explored, our data display a
significant decrease in \bluefrac between high and low
redshifts. Extrapolating the observed trends further back in time up
to $z \sim 1$, one can speculate that there must have been a time when
most galaxies resided in the blue cloud, irrespective of their
environment.  Panel (a) of Fig.~\ref{fig:fbz_mass} clearly suggests
that the time when blue galaxies were in the majority has ended
earlier for galaxies in groups than for those in field or isolated
galaxies, and a similar trend is present, albeit at a far lower
significance, also for galaxies in panels (b) and (c).

Unfortunately, as shown by Table~\ref{tab:massvollimfit}, the error
bars in the slopes of the fit are quite large and it is difficult to
draw definitive conclusions on the fractional rate of change in
\bluefrac for the different environments of each mass bin considered.
In each redshift bin, all the values obtained for $\beta$ are
compatible with each other for the three environments considered,
given their large error bars.

In parallel with the analysis completed in Section
\ref{sect:VolLimSamples} for the mass volume-limited samples we
proceeded to plot in Fig.~\ref{fig:fbz_mass} results for galaxies
residing in groups homogeneous in richness across the different
panels. We again used the sample of groups with two or more members in
the brighter absolute magnitude cut-off sample, and then repeated our
measurements of \bluefrac for the group members satisfying the
mass-bins limits. The numbers of the galaxy group samples defined in
this way are those indicated by the entry Group galaxies II in
Table~\ref{tab:massbinnumb}.

With the possible exception of panel (a), moving to richer groups does
not seem to affect significantly the value of $\it{F_{blue}}$. This result
is consistent with the previous one: only for galaxies of lower
stellar masses do we still see color segregation as a function of
environment and therefore only for these masses can we expect to see a
significant dependence of \bluefrac on group richness.

\begin{table*}
\caption{Summary of fits results for mass bins of Table
~\ref{tab:massbinnumb}.  We parametrized the evolution of \bluefrac
with a fit of the form \bluefrac $ \propto (1+z)^{\beta}$.
{\it Mass} is in units of $log({\cal M}_*/({\cal M}_{\odot} \times h_{70}^{-2}))$.}
\label{tab:massvollimfit}  
\centering                  
\begin{tabular}{c c c c c c c c c }    
\hline\hline  
          & \multicolumn{2}{c}{Sample MM-I}  & \multicolumn{2}{c}{Sample MM-II} & \multicolumn{2}{c}{Sample MM-III} & \multicolumn{2}{c}{Sample MM-IV}  \\   
 Stellar mass range & \multicolumn{2}{c} {$10.0\leq{\it Mass}\leq 10.5$} & \multicolumn{2}{c} {$10.3\leq{\it Mass}\leq 10.8$} & \multicolumn{2}{c} {$10.6\leq{\it Mass}\leq 11.1$}  & \multicolumn{2}{c} {$10.9\leq{\it Mass}\leq 11.4$} \\
    $z$ range   & \multicolumn{2}{c}{$0.1\leq z\leq0.45$} & \multicolumn{2}{c}{$0.1\leq z\leq0.6$} & \multicolumn{2}{c}{$0.1\leq z\leq0.8$} & \multicolumn{2}{c}{$0.1\leq z\leq 1.0$} \\                  
\hline  \hline                    
                   & \bluefrac($z=0)$  & $\beta$ & \bluefrac($z=0)$ & $\beta$ & \bluefrac($z=0)$ & $\beta$ & \bluefrac($z=0)$ & $\beta$ \\
All galaxies       & 0.27$\pm$0.10 & 1.9$\pm$1.2 & 0.22$\pm$0.06 & 1.3$\pm$0.8 & 0.12$\pm$0.04 & 1.4$\pm$0.7 & 0.06$\pm$0.04 &  1.0$\pm$1.2 \\
Isolated galaxies      & 0.32$\pm$0.18 & 2.4$\pm$1.7 & 0.11$\pm$0.06 & 3.7$\pm$1.4 & 0.11$\pm$0.06 & 2.3$\pm$1.1 & 0.14$\pm$0.20 & -0.5$\pm$2.6 \\
Group galaxies I   & 0.15$\pm$0.11 & 2.7$\pm$2.4 & 0.16$\pm$0.08 & 1.5$\pm$1.5 & 0.10$\pm$0.05 & 1.6$\pm$1.1 & 0.10$\pm$0.07 & -0.3$\pm$1.5 \\
Group galaxies II  & 0.13$\pm$1.18 & 2.0$\pm$9.9 & 0.20$\pm$0.19 & 0.6$\pm$3.1 & 0.12$\pm$0.08 & 1.2$\pm$1.3 & 0.10$\pm$0.06 & -0.3$\pm$1.5 \\
\hline                             
\end{tabular}
\end{table*}

An interesting trend suggested by Fig.~\ref{fig:fbz_mass} is that for
more massive galaxies the predominance of the redder-color galaxy
population started earlier in cosmic time than for lower mass
galaxies. We decided to investigate this trend directly by plotting at
fixed redshift \bluefrac as a function of mass.

\begin{figure*}
\centering 
\includegraphics[width=10cm,angle=0,angle=270]{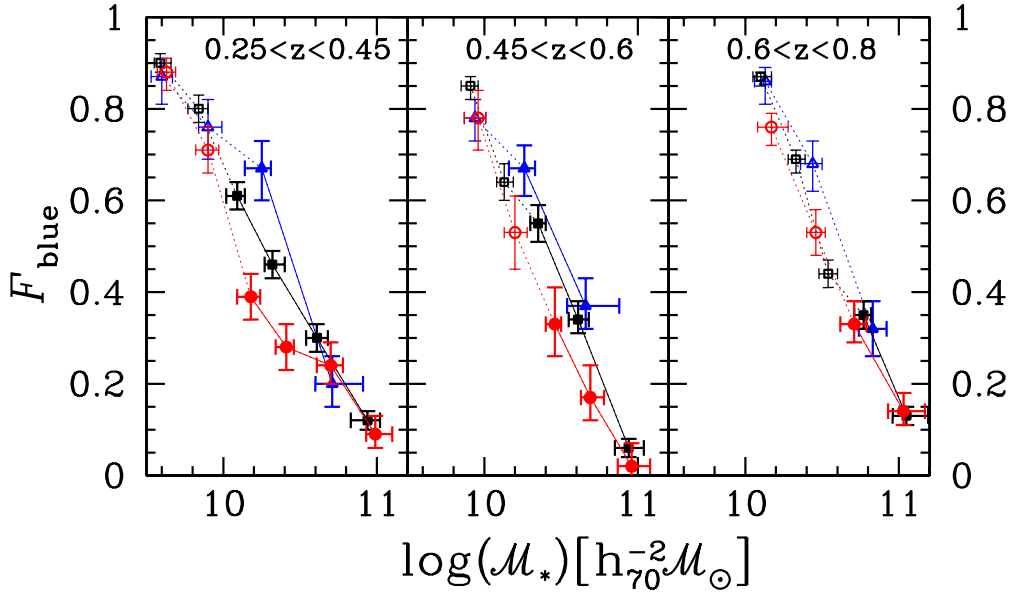}
\caption{The three panels show, in the redshift bin indicated by the
top labels, \bluefrac as a function of galaxy stellar mass. Red
circles refer to group galaxies, blue triangles to isolated galaxies,
while black squares to the total galaxy population. Error bars along
the y-axis are those obtained by bootstrap, and those along the x-axis
indicates the inter-quartiles ranges of the mass distribution in the
bin under scrutiny.}
\label{fig:fixedz}
\end{figure*}

The results are shown in Fig.~\ref{fig:fixedz}.
The label on each of the three panels shows the redshift ranges
adopted, while the color code is, as usual, blue, red, and black for
isolated, group and field galaxies, respectively. 

We adopted the following three redshift bins where to perform this
analysis [0.25:0.45],[0.45:0.60] and [0.60:0.80].  For these bins, we
had already defined complete mass limited samples as listed in
Table~\ref{tab:masslimnumb}. However, to increase the range of masses
probed in each redshift bin we decided to extend the analysis down to
masses where, according to the procedure described in
Section~\ref{sect:masscompl}, we had in the redshift bin considered a
completeness lower than $85\%$ for the reddest galaxies of our
sample. Obviously such a strategy can be adopted only if one is sure
that a representative (in color) sample of the lower mass galaxies
under scrutiny is observed in a large fraction of the volume under
consideration, so that it can be statistically reconstructed, \eg
applying a correction using the $V/V_{max}$ technique. We therefore
lowered our mass limit to masses such that the completeness even for
the reddest galaxies was always around $100\%$ at the lowest limit of
the redshift bin considered. We then weighted each observed galaxy
with its corresponding volume correction, estimated as the ratio of
the volume contained within the [$z_{min}$:$z_{max}$] bin and the
actual volume up to which the galaxy can be observed within the survey
$I_{AB} \leq 22.5$ selection.

Filled points in Fig.~\ref{fig:fixedz} are those referring to bins
in masses where we are complete, while empty points refer to the lower
masses bins where the $V/V_{max}$ corrections discussed above have
been applied.

In the x-axis we plot the median mass value for each mass bin, while
in the y-axis is the value of \bluefrac for the same bin.  The
error-bars along the y-axis are those obtained with bootstrap analysis,
and the ones along the x-axis indicate the inter-quartiles ranges of
the mass distribution within the mass bin considered.  The choice of
the mass bins is somewhat arbitrary, as for the isolated galaxies
sample we were forced to split the sample in fewer bins for
statistical reasons.

The trends displayed in the lowest redshift bin well agree
qualitatively with similar trends detected at $z \sim 0$
\citep[\eg][]{Kauffmann2004, Baldry2006}, showing a clear dependence
of \bluefrac both on mass and on environment. However, there are some
new interesting points, obserbed thanks to the unprecedented wide
redshift/mass ranges covered by our dataset.

On one hand, at all redshift bins more massive galaxies always display
a lower $\it{F_{blue}}$, near to zero values, irrespective of the
environment they live in, while for lower mass galaxies the value of
\bluefrac raises towards unity, again irrespective of the environment
they live in. Therefore, in each redshift range, and more clearly in
the first two ones plotted, where the mass coverage is wider, we
witness the presence of a progressive saturation of \bluefrac towards
high/low values at the extremes of the mass ranges studied. However,
there is a restricted range of masses for which the color of galaxies
show a visible dependence on environment. This mass range is the one
where both sides of the bimodal distribution of galaxy colors are well
populated and we can detect a considerable environment dependent
variation of $\it{F_{blue}}$.  This result echoes a similar one obtained by
\citet{Kauffmann2004} in the local Universe.

On the other hand, moving from lower to higher redshifts we witness a
progressive increase of \bluefrac for each mass bin, with the possible
exception of the highest masses, as already observed in the previous
section.  Such decrease of \bluefrac as cosmic time goes by seem to be
accompanied by a progressive {\it opening} along the x-axis of the
difference between the different environments, most prominent in the
mass ranges for which \bluefrac $\sim 0.5$.

\subsection{Detection of the possible signature of environmental effects}  
\label{sect:signature}   

It is interesting to quantify the trend discussed at the end of the
previous section using a simple parameter: the value, for each mass
bin and environment, of the redshift when \bluefrac $= 0.5$. We can
call this quantity {\it t$_{50-50}$} to indicate that it corresponds
to the time when the galaxies in the environment and mass bin
considered were equally partitioned between blue and red colors.
Although obtained through a slightly different type of analysis, this
quantity is equivalent to the transitional mass m$_{tr}$ identified by
various authors, both in the low redshift regime by \citet{Baldry2004}
and \citet{Kauffmann2004} and at higher redshifts by \citet{Bundy2006}
and for the \10K by Bolzonella et al. (2009).

\begin{figure*}
\sidecaption 
\includegraphics[width=12cm,angle=0]{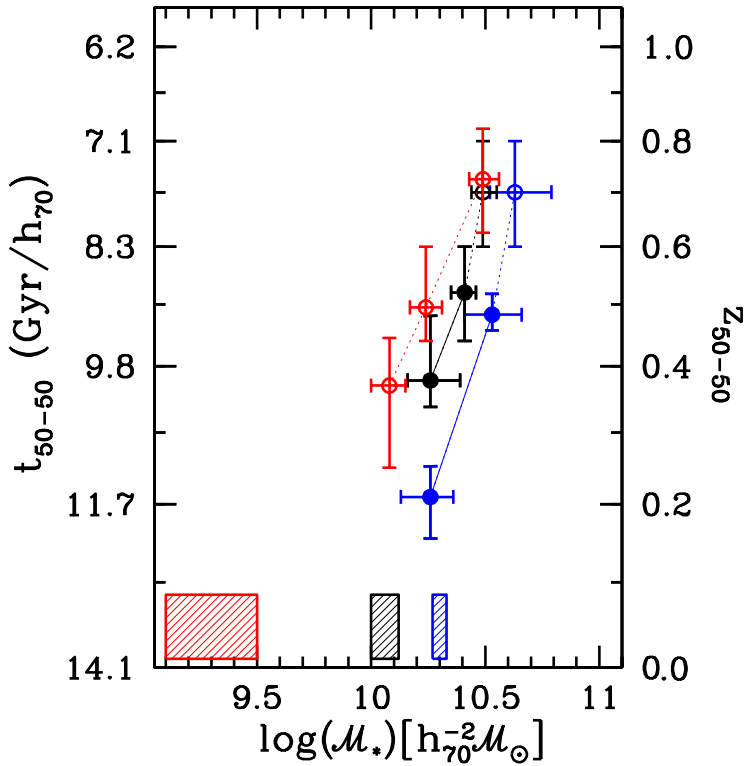} 
\caption{The time {\it t$_{50-50}$}, at which \bluefrac$ \sim 0.5$, is
plotted as a function of galaxy stellar mass. The left scale is in
units of cosmic time, in Gyrs, while the right scale is in
redshift. The triangles, circles, and squares refer to the sample of
isolated, group and field galaxies, respectively.  Filled points are
those corresponding to bins in masses where we are complete, while
empty points refer to the lower mass bins, where the $V/V_{max}$
corrections are needed. The shaded boxes have been obtained from
\citet{Baldry2006}. See text for more details.}
\label{fig:down}
\end{figure*}

Figure~\ref{fig:down} shows the value of {\it t$_{50-50}$}, expressed
in units of time on the left-hand scale and redshift on the right-hand
scale, for different galaxy stellar masses. The triangles, circles,
and squares refer to the sample of isolated, group and field galaxies,
respectively. Filled points are estimated from Fig.~\ref{fig:fbz_mass}
using the fits to the points plotted in each mass bin as shown in
Table~\ref{tab:massvollimfit}. The values obtained in this way for
{\it t$_{50-50}$} do not need any incompleteness correction, since
they are observed directly in our \10K in a mass range where we are
complete, but they cover only a limited range of mass and
environments. Empty points are obtained using a $V/V_{max}$ correction
and Fig.~\ref{fig:fixedz}. These values for {\it t$_{50-50}$} are
therefore more uncertain, since they are based on incompleteness
corrections.

For filled points, the error bars along the x-axis link the upper and
lower quartiles of the mass distribution in the mass bin considered,
while the error bars along the y-axis are obtained by the {\it r.m.s.}
of the value of {\it t$_{50-50}$} as obtained by bootstrapping the
sample of galaxies that enters in the corresponding panels of
Fig.~\ref{fig:fbz_mass}. For empty points, the error bars along the
y-axis indicate the redshift bin where the value of {\it t$_{50-50}$} was
estimated, while the error bars along the x-axis show the upper and
lower interval for the masses as obtained from the error bars in
Fig.~\ref{fig:fixedz}.

For the sake of comparison, we have added boxes indicating masses for
which {\it t$_{50-50}$} $\sim 0.05$, as obtained from the relationship
between mass, galaxy colors, and environment in the local Universe
determined in \citet{Baldry2006}. These values are only indicative and
extracted from their Fig. 11, panel b, where curves of the fraction of
red galaxies \vs stellar mass are shown in 12 different bins of galaxy
density.  We choose to use as representative of the global population
the two central curves of their plot, corresponding to $-0.2 \leq
log(\Sigma) \leq 0.2$, while for isolated/group galaxies in the local
Universe we used the curves covering densities $-0.8 \leq log(\Sigma)
\leq -0.4$ and $+0.4 \leq log(\Sigma) \leq +0.8$. This last choice was
made considering that our total span in densities is not as wide as
theirs: the bulk of the galaxy population in our groups/isolated
galaxies is located in regions that are roughly a factor of 3
above/below median densities (see Fig. 15, panel a in Kovac et al.,
2009 and a similar plot for our isolated galaxy population).

The vertical size of the boxes plotted in Fig.~\ref{fig:down}
corresponds to the redshift range of the sample used in
\citet{Baldry2006}, while the mass range corresponding to the
horizontal box size corresponds to the range of galaxy stellar mass
values where the fraction of galaxies in the red sequence equals 0.5
for the three different environments defined by the curves mentioned
above. It should be noticed that the masses in \citet{Baldry2006} were
estimated using a \citet{Kroupa2001} initial mass function and
recurrent bursts of star formation superimposed to continuous models
of star formation. As a consequence the \citet{Baldry2006} masses have
a systematically offset towards higher values with respect to our
masses by a factor that can be as high as 0.15 dex, possibly
explaining the slight offset of local points with respect to the
trends displayed by our high-z points.
 
Figure~\ref{fig:down} highlights the main result of our paper. The
first visible trend is that as cosmic time goes by the typical mass at
which the galaxy population is equally partitioned between red and
blue galaxies moves progressively to lower values. This is another way
of expressing the well-known downsizing pattern observed in galaxy
population evolution. In our plot the decrease in galaxy K-band
luminosity with decreasing redshift of galaxies dominated by star
formation, as originally reported by \citet{Cowie1996}, translates
into a progressive increase of {\it t$_{50-50}$} when considering
galaxies of lower masses.

But this global behavior displays differences depending on the subset
of the galaxy population we are considering. A consistent trend
emerges, despite the large error bars: for each mass considered {\it
t$_{50-50}$} is progressively delayed moving from groups to the field
and to the isolated galaxy population. In other words the downsizing
for the galaxy population is further modulated by the environment:
galaxies located in more massive halos (groups) become red earlier in
cosmic time, a trend that shows again a downsizing behavior on the
larger scales now considered. The trends displayed by our data well
match those observed in the local Universe by \citet{Baldry2006}.

Last but not least, another interesting trend suggested by
Fig.~\ref{fig:down} is the convergence, visible at higher masses, of
the value of {\it t$_{50-50}$}, irrespective of the environment
considered. We are aware that this interpretation is plagued by
uncertainties, as for these high masses the redshifts are
correspondingly higher and those most affected by various
incompleteness/contamination effects. Two possible biases can be at
play at higher redshifts: the progressive degradation in the
efficiency of the group/isolated galaxies algorithms and the
progressive incompleteness towards the red galaxy population. Both
biases act in the direction of reducing the differences between the
color properties of the galactic populations we are studying. However
we expect that these two biases are minimal, as discussed at length in
the previous sections.

One should also consider that for galaxies residing in more extreme
density regimes, as those represented by rich cluster cores and not
observed in our sample, there could be a residual difference even at
redshift $\sim 1$ from the general galaxy population \citep[see, \eg][]{Tanaka2008}. As already suggested in the introduction, however,
the physical mechanisms responsible for these differences are
presumably not the same as those at play in the group environment.
Our results parallel those obtained, although by a different kind of
analysis, by Bolzonella et al. (2009), and those shown for galaxies
morphologies by \citet{Kovac2009b}.  

We note that the evidence we presented, \ie the faster shutdown of
star formation in group environment - as the color transition from
blue to red galaxy can broadly be interpreted - cannot be interpreted
utilizing only {\it ab initio}/internal mechanisms, and was obtained
thanks to the unprecedented wide redshift, mass and environment ranges
covered by our survey.

The different value of {\it t$_{50-50}$} for galaxies of different
stellar masses, irrespective of environment, can be explained by
resorting to internally driven mechanisms shutting-down star
formation. The presence of an active galactic nucleus (AGN) feedback
and shock-heating physics can be enough to explain the
anti--hierarchical nature of the relation between stellar mass and
stellar age of galaxies, because these mechanisms can be more
efficient in more massive galaxies \citep[see][]{Birnboim2003,
Bower2006, Bundy2006, Croton2006, Cattaneo2008}.

In a similar way, the detection of an offset in the value of {\it
t$_{50-50}$} between samples of group/isolated galaxies at fixed
stellar mass does not necessarily imply that nurture mechanisms are at
work. It could be explained by a different time of assembly of
galaxies in haloes of different masses, a {\it nature} mechanism that
results in a more evolved galaxy population in groups and clusters, at
fixed stellar mass, in a given redshift bin \citep[see][]{Gao2005,
Balogh2007}.

In contrast, the trend suggested by Fig.~\ref{fig:down} indicates that
the migration of galaxies from the blue cloud to the red sequence is a
process more efficient/faster in groups than in isolated/field
galaxies, and is therefore the signature of environmental processes at
play in groups in shaping galaxy evolution.

Interestingly, such mechanisms seem to become progressively more
relevant moving to lower galaxy stellar masses, while they seem to be
irrelevant to galaxies of higher stellar masses, at least in the
redshift range we explored (see also Bolzonella et al., 2009).  We can
therefore distinguish between two different channels for the
production of red galaxies, corresponding respectively, to use a
common nomenclature, to {\it nature} red galaxies and {\it nurture}
red galaxies. Our results suggest that galaxies of masses $\approx
10.8$ solar in logarithmic scale are already in place by $z \sim 1$
and their origin could be due primarily to so-called {\it
nature}/internal mechanisms, as no strong environmental dependency is
visible up to $z \sim 1$.  In contrast, for masses below this value
and at redshifts lower than $z\sim 1$, we witness the emergence in
groups of an additional contribution of red galaxies. This is what we
can call {\it nurture} red galaxies: galaxies that deviate slightly
from the trend of the downsizing scenario as displayed by the global
galaxy population. This nurture population is the one responsible for
the earlier value of {\it t$_{50-50}$} in groups, and its importance
grows as cosmic time goes by, causing the steady growth in the
difference of {\it t$_{50-50}$} moving to lower galaxy masses.

There are various mechanisms that occur in groups and that are more
efficient for less massive galaxies, including the gradual cessation
of star formation induced by gentle gas stripping and starvation by a
diffuse intragroup medium, or by slow group-scale harassment
\citep[see, \eg][]{Larson1980, Moore1999, Gnedin2003, Roediger2005,
Kawata2008}. These mechanisms could be natural candidates for
explaining the trends we observe. Their increasing importance after
$z\sim1$ most probably mirrors the progressive emergence of
structures, as predicted by hierarchical clustering growth scenario,
where such mechanisms can effectively take place.

\section{Summary and conclusions}   
\label{sect:Concl} 

Taking advantage of the large coverage both of redshift and
galaxy/group properties in the 10K galaxy/groups catalogue, we
revisited of the blueing of the galaxy population in groups toward
higher redshift, originally observed by Butcher and Oemler (1978),
gaining some interesting new insights, that can be summarized as
follows.

1. We have showed that using rest-frame B-band volume-limited samples,
the group galaxy population becomes bluer as redshift increases, but
maintains a systematic difference with respect to the global galaxy
population, and an even larger difference with respect to the isolated
galaxy population. Superimposed on this global effect, we detected
additional trends as a function of both galaxy B-band rest-frame
luminosity and group properties. More luminous galaxies exhibit
stronger variations in \bluefrac among group, field, and isolated
environments and groups richer or with higher $\sigma$ show a lower
$\it{F_{blue}}$.

2. The difference between the three different environments
increases between high and low redshift. At the highest redshift bin
explored ($z \sim 1$), there is a small but still significant
difference in \bluefrac among group,field, and isolated samples.
This gradual increase in the \bluefrac difference with cosmic time is
a clear signature of an environmental dependence, but not necessarily
of the existence of environmental effects at work. It could be the
result of an {\it ab initio} bias that favors later formation of
lower-mass galaxies in lower density environments, causing the delayed
and more efficient replenishing of the blue cloud in lower density
environments.

3. Moving to mass-selected samples, a necessary step in clarifying the
key mechanisms in determining the relationship observed using
luminosity-selected samples, allows us to realize almost immediately
that at least part of the strong trends observed when using rest-frame
evolving B-band volume-limited samples are caused by the large
population of lower mass, bright blue galaxies for which we miss the
redder, equally low mass, counterparts. In other words, the biased
view imposed by the B-band luminosity selection amplifies the findings
obtained using B-band volume-limited samples.

4. Another effect has to be taken into consideration if one wants to
disentangle the mass and environment influence on galaxy colors. The
existence of different mass functions in different environments (see
Bolzonella et al. 2009) forces us to work in mass bins narrow enough
so that any color segregation cannot be attributed simply to the
different mass distribution.

5. The first outcome of this careful analysis is that there is still a
significant residual difference in color as a function of environment
only for the lowest mass bin explored (${\it Mass}\leq 10.6 $, solar
masses in logarithmic scale), while this difference progressively
disappears moving to higher masses.

6. By using a $V/V_{max}$ correction, we can extend our analysis to
lower masses, witnessing, in all the redshift range we explore, the
presence of progressive saturation of \bluefrac towards high/low
values at the extremes of the mass ranges studied. At each redshift,
there is a restricted range of masses for which the color of galaxies
show a visible dependence on environment, and as cosmic time increases
the typical mass at which the galaxy population is equally partitioned
between red and blue galaxies moves progressively to lower values.
This pattern, consistent with the well known downsizing pattern
observed in galaxy population evolution, is further modulated by
environment: galaxies located in more massive halos (groups) become
red earlier in cosmic time.

7. Finally our most interesting finding is that there is evidence that
the color transition from blue to red galaxies seems to be faster in
groups as cosmic time increases. In other words, we seem to witness
the slow emergence of an environmental/nurture effect on galaxy
evolution, which causes the faster migration of galaxies from the blue
cloud to the red sequence in groups (with respect to isolated/field
galaxies) and effect that becomes more relevant moving from higher to
lower galaxy stellar masses \citep[see also][for a parallel analysis
using galaxy morphologies]{Kovac2009b}.

8. Our results suggest that galaxies of ${\it Mass} \approx 10.8$
solar masses in logarithmic scale are already in place by $z \sim 1$
and their origin could be due primarily to so-called {\it
  nature}/internal mechanisms, since no strong environmental
dependency is detectable up to $z \sim 1$.

9. In contrast, for masses below this value and at redshifts lower
than $z\sim 1$ we witness the emergence of what we call {\it nurture}
red galaxies: galaxies that slightly deviate from the trend of the
downsizing scenario displayed by the global galaxy population and do
more so as cosmic time progresses.

There are various mechanisms that occur in groups and are more
efficient for less massive galaxies (gradual cessation of star
formation induced by gentle gas stripping and starvation by a diffuse
intragroup medium, or by slow group-scale harassment). These
mechanisms could be the natural candidates to explain the trends
observed after $z\sim1$, a timing that could simply mirror that of
the progressive emergence of structures where these mechanisms can
effectively take place.

The completion of zCOSMOS bright, and subsequent availability of \20K,
will enable us to place on a more robust basis this result, which
indicates that environment starts playing an active role in shaping
galaxy evolution after $z \sim 1$.

\begin{acknowledgements}
We acknowledge support from an INAF contract PRIN-2007/1.06.10.08 and 
an ASI grant ASI/COFIS/WP3110 I/026/07/0.
\end{acknowledgements}

\bibliographystyle{aa}
\bibliography{biblio}

\begin{thebibliography}{111}
\expandafter\ifx\csname natexlab\endcsname\relax\def\natexlab#1{#1}\fi

\bibitem[{{Abell}(1958)}]{Abell1958}
{Abell}, G.~O. 1958, \apjs, 3, 211

\bibitem[{{Allington-Smith} {et~al.}(1993){Allington-Smith}, {Ellis}, {Zirbel},
  \& {Oemler}}]{Allington-Smith1993}
{Allington-Smith}, J.~R., {Ellis}, R., {Zirbel}, E.~L., \& {Oemler}, A.~J.
  1993, \apj, 404, 521

\bibitem[{{Andreon} \& {Ettori}(1999)}]{Andreon1999}
{Andreon}, S. \& {Ettori}, S. 1999, \apj, 516, 647

\bibitem[{{Andreon} {et~al.}(2004){Andreon}, {Lobo}, \& {Iovino}}]{Andreon2004}
{Andreon}, S., {Lobo}, C., \& {Iovino}, A. 2004, \mnras, 349, 889

\bibitem[{{Andreon} {et~al.}(2006){Andreon}, {Quintana}, {Tajer}, {Galaz}, \&
  {Surdej}}]{Andreon2006}
{Andreon}, S., {Quintana}, H., {Tajer}, M., {Galaz}, G., \& {Surdej}, J. 2006,
  \mnras, 365, 915

\bibitem[{{Baldry} {et~al.}(2006){Baldry}, {Balogh}, {Bower}, {Glazebrook},
  {Nichol}, {Bamford}, \& {Budavari}}]{Baldry2006}
{Baldry}, I.~K., {Balogh}, M.~L., {Bower}, R.~G., {et~al.} 2006, \mnras, 373,
  469

\bibitem[{{Baldry} {et~al.}(2004){Baldry}, {Glazebrook}, {Brinkmann},
  {Ivezi{\'c}}, {Lupton}, {Nichol}, \& {Szalay}}]{Baldry2004}
{Baldry}, I.~K., {Glazebrook}, K., {Brinkmann}, J., {et~al.} 2004, \apj, 600,
  681

\bibitem[{{Balogh} {et~al.}(2004{\natexlab{a}}){Balogh}, {Eke}, {Miller},
  {Lewis}, {Bower}, {Couch}, {Nichol}, {Bland-Hawthorn}, {Baldry}, {Baugh},
  {Bridges}, {Cannon}, {Cole}, {Colless}, {Collins}, {Cross}, {Dalton}, {de
  Propris}, {Driver}, {Efstathiou}, {Ellis}, {Frenk}, {Glazebrook}, {Gomez},
  {Gray}, {Hawkins}, {Jackson}, {Lahav}, {Lumsden}, {Maddox}, {Madgwick},
  {Norberg}, {Peacock}, {Percival}, {Peterson}, {Sutherland}, \&
  {Taylor}}]{Balogh2004b}
{Balogh}, M., {Eke}, V., {Miller}, C., {et~al.} 2004{\natexlab{a}}, \mnras,
  348, 1355

\bibitem[{{Balogh} {et~al.}(2004{\natexlab{b}}){Balogh}, {Baldry}, {Nichol},
  {Miller}, {Bower}, \& {Glazebrook}}]{Balogh2004a}
{Balogh}, M.~L., {Baldry}, I.~K., {Nichol}, R., {et~al.} 2004{\natexlab{b}},
  \apjl, 615, L101

\bibitem[{{Balogh} {et~al.}(2007){Balogh}, {Wilman}, {Henderson}, {Bower},
  {Gilbank}, {Whitaker}, {Morris}, {Hau}, {Mulchaey}, {Oemler}, \&
  {Carlberg}}]{Balogh2007}
{Balogh}, M.~L., {Wilman}, D., {Henderson}, R.~D.~E., {et~al.} 2007, \mnras,
  374, 1169

\bibitem[{{Beers} {et~al.}(1990){Beers}, {Flynn}, \& {Gebhardt}}]{Beers1990}
{Beers}, T., {Flynn}, K., \& {Gebhardt}, K. 1990, \apj, 100, 32

\bibitem[{{Berlind} {et~al.}(2006){Berlind}, {Frieman}, {Weinberg}, {Blanton},
  {Warren}, {Abazajian}, {Scranton}, {Hogg}, {Scoccimarro}, {Bahcall},
  {Brinkmann}, {Gott}, {Kleinman}, {Krzesinski}, {Lee}, {Miller}, {Nitta},
  {Schneider}, {Tucker}, \& {Zehavi}}]{Berlind2006}
{Berlind}, A.~A., {Frieman}, J., {Weinberg}, D.~H., {et~al.} 2006, \apjs, 167,
  1

\bibitem[{{Birnboim} \& {Dekel}(2003)}]{Birnboim2003}
{Birnboim}, Y. \& {Dekel}, A. 2003, \mnras, 345, 349

\bibitem[{{Biviano} {et~al.}(1997){Biviano}, {Katgert}, {Mazure}, {Moles}, {den
  Hartog}, {Perea}, \& {Focardi}}]{Biviano1997}
{Biviano}, A., {Katgert}, P., {Mazure}, A., {et~al.} 1997, \aap, 321, 84

\bibitem[{{Blanton} \& {Berlind}(2007)}]{BlantonBerlind2007}
{Blanton}, M.~R. \& {Berlind}, A.~A. 2007, \apj, 664, 791

\bibitem[{{Blanton} {et~al.}(2006){Blanton}, {Eisenstein}, {Hogg}, \&
  {Zehavi}}]{Blanton2006a}
{Blanton}, M.~R., {Eisenstein}, D., {Hogg}, D.~W., \& {Zehavi}, I. 2006, \apj,
  645, 977

\bibitem[{{Bolzonella} {et~al.}(2009){Bolzonella}, {Kova{\v c}}, {Pozzetti},
  {Zucca}, {Cucciati}, {Lilly}, {Peng}, {Iovino}, {Zamorani}, {Vergani},
  {Tasca}, {Lamareille}, {Oesch}, {Caputi}, {Kampczyk}, {Bardelli}, {Maier},
  {Abbas}, {Knobel}, {Scodeggio}, {Carollo}, {Contini}, {Kneib}, {Le Fevre},
  {Mainieri}, {Renzini}, {Bongiorno}, {Coppa}, {de la Torre}, {de Ravel},
  {Franzetti}, {Garilli}, {Le Borgne}, {Le Brun}, {Mignoli}, {Pello},
  {Perez-Montero}, {Ricciardelli}, {Silverman}, {Tanaka}, {Tresse}, {Bottini},
  {Cappi}, {Cassata}, {Cimatti}, {Guzzo}, {Koekemoer}, {Leauthaud}, {Maccagni},
  {Marinoni}, {McCracken}, {Memeo}, {Meneux}, {Porciani}, {Scaramella},
  {Aussel}, {Capak}, {Ilbert}, {Kartaltepe}, {Salvato}, {Sanders}, {Scarlata},
  {Scoville}, {Taniguchi}, \& {Thompson}}]{Bolzonella2009}
{Bolzonella}, M., {Kova{\v c}}, K., {Pozzetti}, L., {et~al.} 2009, ArXiv
  e-prints 0907.0013

\bibitem[{{Bottini} {et~al.}(2005){Bottini}, {Garilli}, {Maccagni}, {Tresse},
  {Le Brun}, {Le F{\`e}vre}, {Picat}, {Scaramella}, {Scodeggio}, {Vettolani},
  {Zanichelli}, {Adami}, {Arnaboldi}, {Arnouts}, {Bardelli}, {Bolzonella},
  {Cappi}, {Charlot}, {Ciliegi}, {Contini}, {Foucaud}, {Franzetti}, {Guzzo},
  {Ilbert}, {Iovino}, {McCracken}, {Marano}, {Marinoni}, {Mathez}, {Mazure},
  {Meneux}, {Merighi}, {Paltani}, {Pollo}, {Pozzetti}, {Radovich}, {Zamorani},
  \& {Zucca}}]{Bottini2005}
{Bottini}, D., {Garilli}, B., {Maccagni}, D., {et~al.} 2005, \pasp, 117, 996

\bibitem[{{Bower} {et~al.}(2006){Bower}, {Benson}, {Malbon}, {Helly}, {Frenk},
  {Baugh}, {Cole}, \& {Lacey}}]{Bower2006}
{Bower}, R.~G., {Benson}, A.~J., {Malbon}, R., {et~al.} 2006, \mnras, 370, 645

\bibitem[{{Br{\"u}ggen} \& {De Lucia}(2008)}]{Bruggen2008}
{Br{\"u}ggen}, M. \& {De Lucia}, G. 2008, \mnras, 383, 1336

\bibitem[{{Bruzual} \& {Charlot}(2003)}]{BruzualCharlot2003}
{Bruzual}, G. \& {Charlot}, S. 2003, \mnras, 344, 1000

\bibitem[{{Bundy} {et~al.}(2006){Bundy}, {Ellis}, {Conselice}, {Taylor},
  {Cooper}, {Willmer}, {Weiner}, {Coil}, {Noeske}, \& {Eisenhardt}}]{Bundy2006}
{Bundy}, K., {Ellis}, R.~S., {Conselice}, C.~J., {et~al.} 2006, \apj, 651, 120

\bibitem[{{Butcher} \& {Oemler}(1978)}]{ButcherOemler1978}
{Butcher}, H. \& {Oemler}, Jr., A. 1978, \apj, 219, 18

\bibitem[{{Butcher} \& {Oemler}(1984)}]{ButcherOemler1984}
{Butcher}, H. \& {Oemler}, Jr., A. 1984, \apj, 285, 426

\bibitem[{{Cattaneo} {et~al.}(2008){Cattaneo}, {Dekel}, {Faber}, \&
  {Guiderdoni}}]{Cattaneo2008}
{Cattaneo}, A., {Dekel}, A., {Faber}, S.~M., \& {Guiderdoni}, B. 2008, \mnras,
  389, 567

\bibitem[{{Chabrier}(2003)}]{Chabrier2003}
{Chabrier}, G. 2003, \apjl, 586, L133

\bibitem[{{Coleman} {et~al.}(1980){Coleman}, {Wu}, \& {Weedman}}]{Coleman1980}
{Coleman}, G.~D., {Wu}, C.-C., \& {Weedman}, D.~W. 1980, \apjs, 43, 393

\bibitem[{{Cooper} {et~al.}(2007){Cooper}, {Newman}, {Coil}, {Croton}, {Gerke},
  {Yan}, {Davis}, {Faber}, {Guhathakurta}, {Koo}, {Weiner}, \&
  {Willmer}}]{Cooper2007}
{Cooper}, M.~C., {Newman}, J.~A., {Coil}, A.~L., {et~al.} 2007, \mnras, 376,
  1445

\bibitem[{{Cooper} {et~al.}(2006){Cooper}, {Newman}, {Croton}, {Weiner},
  {Willmer}, {Gerke}, {Madgwick}, {Faber}, {Davis}, {Coil}, {Finkbeiner},
  {Guhathakurta}, \& {Koo}}]{Cooper2006}
{Cooper}, M.~C., {Newman}, J.~A., {Croton}, D.~J., {et~al.} 2006, \mnras, 370,
  198

\bibitem[{{Cooray} \& {Sheth}(2002)}]{CooraySheth2002}
{Cooray}, A. \& {Sheth}, R. 2002, \physrep, 372, 1

\bibitem[{{Cowie} {et~al.}(1996){Cowie}, {Songaila}, {Hu}, \&
  {Cohen}}]{Cowie1996}
{Cowie}, L.~L., {Songaila}, A., {Hu}, E.~M., \& {Cohen}, J.~G. 1996, \aj, 112,
  839

\bibitem[{{Croton} {et~al.}(2006){Croton}, {Springel}, {White}, {De Lucia},
  {Frenk}, {Gao}, {Jenkins}, {Kauffmann}, {Navarro}, \& {Yoshida}}]{Croton2006}
{Croton}, D.~J., {Springel}, V., {White}, S.~D.~M., {et~al.} 2006, \mnras, 365,
  11

\bibitem[{{Cucciati} {et~al.}(2009{\natexlab{a}}){Cucciati}, {Iovino}, {Kova{\v
  c}}, {Pinco}, {Pallo}, {Pinco}, \& {Pallo}}]{Cucciati2009b}
{Cucciati}, O., {Iovino}, A., {Kova{\v c}}, K., {et~al.} 2009{\natexlab{a}},
  \aap, submitted

\bibitem[{{Cucciati} {et~al.}(2006){Cucciati}, {Iovino}, {Marinoni}, {Ilbert},
  {Bardelli}, {Franzetti}, {Le F{\`e}vre}, {Pollo}, {Zamorani}, {Cappi},
  {Guzzo}, {McCracken}, {Meneux}, {Scaramella}, {Scodeggio}, {Tresse}, {Zucca},
  {Bottini}, {Garilli}, {Le Brun}, {Maccagni}, {Picat}, {Vettolani},
  {Zanichelli}, {Adami}, {Arnaboldi}, {Arnouts}, {Bolzonella}, {Charlot},
  {Ciliegi}, {Contini}, {Foucaud}, {Gavignaud}, {Marano}, {Mazure}, {Merighi},
  {Paltani}, {Pell{\`o}}, {Pozzetti}, {Radovich}, {Bondi}, {Bongiorno},
  {Busarello}, {de La Torre}, {Gregorini}, {Lamareille}, {Mathez}, {Mellier},
  {Merluzzi}, {Ripepi}, {Rizzo}, {Temporin}, \& {Vergani}}]{Cucciati2006}
{Cucciati}, O., {Iovino}, A., {Marinoni}, C., {et~al.} 2006, \aap, 458, 39

\bibitem[{{Cucciati} {et~al.}(2009{\natexlab{b}}){Cucciati}, {Marinoni},
  {Iovino}, {Pinco}, {Pallo}, {Pinco}, \& {Pallo}}]{Cucciati2009a}
{Cucciati}, O., {Marinoni}, M., {Iovino}, A., {et~al.} 2009{\natexlab{b}},
  \aap, submitted

\bibitem[{{De Lucia} {et~al.}(2007){De Lucia}, {Poggianti},
  {Arag{\'o}n-Salamanca}, {White}, {Zaritsky}, {Clowe}, {Halliday}, {Jablonka},
  {von der Linden}, {Milvang-Jensen}, {Pell{\'o}}, {Rudnick}, {Saglia}, \&
  {Simard}}]{DeLucia2007}
{De Lucia}, G., {Poggianti}, B.~M., {Arag{\'o}n-Salamanca}, A., {et~al.} 2007,
  \mnras, 374, 809

\bibitem[{{De Lucia} {et~al.}(2006){De Lucia}, {Springel}, {White}, {Croton},
  \& {Kauffmann}}]{DeLucia2006}
{De Lucia}, G., {Springel}, V., {White}, S.~D.~M., {Croton}, D., \&
  {Kauffmann}, G. 2006, \mnras, 366, 499

\bibitem[{{De Propris} {et~al.}(2004){De Propris}, {Colless}, {Peacock},
  {Couch}, {Driver}, {Balogh}, {Baldry}, {Baugh}, {Bland-Hawthorn}, {Bridges},
  {Cannon}, {Cole}, {Collins}, {Cross}, {Dalton}, {Efstathiou}, {Ellis},
  {Frenk}, {Glazebrook}, {Hawkins}, {Jackson}, {Lahav}, {Lewis}, {Lumsden},
  {Maddox}, {Madgwick}, {Norberg}, {Percival}, {Peterson}, {Sutherland}, \&
  {Taylor}}]{DePropris2004}
{De Propris}, R., {Colless}, M., {Peacock}, J.~A., {et~al.} 2004, \mnras, 351,
  125

\bibitem[{{De Propris} {et~al.}(2003){De Propris}, {Stanford}, {Eisenhardt}, \&
  {Dickinson}}]{DePropris2003}
{De Propris}, R., {Stanford}, S.~A., {Eisenhardt}, P.~R., \& {Dickinson}, M.
  2003, \apj, 598, 20

\bibitem[{{Dressler} \& {Gunn}(1982)}]{DresslerGunn1982}
{Dressler}, A. \& {Gunn}, J.~E. 1982, \apj, 263, 533

\bibitem[{{Dressler} \& {Gunn}(1992)}]{DresslerGunn1992}
{Dressler}, A. \& {Gunn}, J.~E. 1992, \apjs, 78, 1

\bibitem[{{Eke} {et~al.}(2004){Eke}, {Baugh}, {Cole}, {Frenk}, {Norberg},
  {Peacock}, {Baldry}, {Bland-Hawthorn}, {Bridges}, {Cannon}, {Colless},
  {Collins}, {Couch}, {Dalton}, {de Propris}, {Driver}, {Efstathiou}, {Ellis},
  {Glazebrook}, {Jackson}, {Lahav}, {Lewis}, {Lumsden}, {Maddox}, {Madgwick},
  {Peterson}, {Sutherland}, \& {Taylor}}]{Eke2004}
{Eke}, V.~R., {Baugh}, C.~M., {Cole}, S., {et~al.} 2004, \mnras, 348, 866

\bibitem[{{Ellingson} {et~al.}(2001){Ellingson}, {Lin}, {Yee}, \&
  {Carlberg}}]{Ellingson2001}
{Ellingson}, E., {Lin}, H., {Yee}, H.~K.~C., \& {Carlberg}, R.~G. 2001, \apj,
  547, 609

\bibitem[{{Fabricant} {et~al.}(1991){Fabricant}, {McClintock}, \&
  {Bautz}}]{Fabricant1991}
{Fabricant}, D.~G., {McClintock}, J.~E., \& {Bautz}, M.~W. 1991, \apj, 381, 33

\bibitem[{{Fairley} {et~al.}(2002){Fairley}, {Jones}, {Wake}, {Collins},
  {Burke}, {Nichol}, \& {Romer}}]{Fairley2002}
{Fairley}, B.~W., {Jones}, L.~R., {Wake}, D.~A., {et~al.} 2002, \mnras, 330,
  755

\bibitem[{{Feldmann} {et~al.}(2006){Feldmann}, {Carollo}, {Porciani}, {Lilly},
  {Capak}, {Taniguchi}, {Le F{\`e}vre}, {Renzini}, {Scoville}, {Ajiki},
  {Aussel}, {Contini}, {McCracken}, {Mobasher}, {Murayama}, {Sanders},
  {Sasaki}, {Scarlata}, {Scodeggio}, {Shioya}, {Silverman}, {Takahashi},
  {Thompson}, \& {Zamorani}}]{Feldmann2006}
{Feldmann}, R., {Carollo}, C.~M., {Porciani}, C., {et~al.} 2006, \mnras, 372,
  565

\bibitem[{{Finoguenov} {et~al.}(2007){Finoguenov}, {Guzzo}, {Hasinger},
  {Scoville}, {Aussel}, {B{\"o}hringer}, {Brusa}, {Capak}, {Cappelluti},
  {Comastri}, {Giodini}, {Griffiths}, {Impey}, {Koekemoer}, {Kneib},
  {Leauthaud}, {Le F{\`e}vre}, {Lilly}, {Mainieri}, {Massey}, {McCracken},
  {Mobasher}, {Murayama}, {Peacock}, {Sakelliou}, {Schinnerer}, {Silverman},
  {Smol{\v c}i{\'c}}, {Taniguchi}, {Tasca}, {Taylor}, {Trump}, \&
  {Zamorani}}]{Finoguenov2007}
{Finoguenov}, A., {Guzzo}, L., {Hasinger}, G., {et~al.} 2007, \apjs, 172, 182

\bibitem[{{Gao} {et~al.}(2005){Gao}, {Springel}, \& {White}}]{Gao2005}
{Gao}, L., {Springel}, V., \& {White}, S.~D.~M. 2005, \mnras, 363, L66

\bibitem[{{Gavazzi} {et~al.}(1996){Gavazzi}, {Pierini}, \&
  {Boselli}}]{Gavazzi1996}
{Gavazzi}, G., {Pierini}, D., \& {Boselli}, A. 1996, \aap, 312, 397

\bibitem[{{Gehrels}(1986)}]{Gehrels1986}
{Gehrels}, N. 1986, \apj, 303, 336

\bibitem[{{Gerke} {et~al.}(2005){Gerke}, {Newman}, {Davis}, {Marinoni}, {Yan},
  {Coil}, {Conroy}, {Cooper}, {Faber}, {Finkbeiner}, {Guhathakurta}, {Kaiser},
  {Koo}, {Phillips}, {Weiner}, \& {Willmer}}]{Gerke2005}
{Gerke}, B.~F., {Newman}, J.~A., {Davis}, M., {et~al.} 2005, \apj, 625, 6

\bibitem[{{Gerke} {et~al.}(2007){Gerke}, {Newman}, {Faber}, {Cooper}, {Croton},
  {Davis}, {Willmer}, {Yan}, {Coil}, {Guhathakurta}, {Koo}, \&
  {Weiner}}]{Gerke2007}
{Gerke}, B.~F., {Newman}, J.~A., {Faber}, S.~M., {et~al.} 2007, \mnras, 376,
  1425

\bibitem[{{Gnedin}(2003)}]{Gnedin2003}
{Gnedin}, O.~Y. 2003, \apj, 589, 752

\bibitem[{{Goto}(2005)}]{Goto2005c}
{Goto}, T. 2005, \mnras, 356, L6

\bibitem[{{Goto} {et~al.}(2003){Goto}, {Okamura}, {Yagi}, {Sheth}, {Bahcall},
  {Zabel}, {Crouch}, {Sekiguchi}, {Annis}, {Bernardi}, {Chong}, {G{\'o}mez},
  {Hansen}, {Kim}, {Knudson}, {McKay}, \& {Miller}}]{Goto2003}
{Goto}, T., {Okamura}, S., {Yagi}, M., {et~al.} 2003, \pasj, 55, 739

\bibitem[{{Hopkins}(2004)}]{Hopkins2004}
{Hopkins}, A.~M. 2004, \apj, 615, 209

\bibitem[{{Huchra} \& {Geller}(1982)}]{HuchraGeller1982}
{Huchra}, J.~P. \& {Geller}, M.~J. 1982, \apj, 257, 423

\bibitem[{{Ilbert} {et~al.}(2009){Ilbert}, {Capak}, {Salvato}, {Aussel},
  {McCracken}, {Sanders}, {Scoville}, {Kartaltepe}, {Arnouts}, {LeFloc'h},
  {Mobasher}, {Taniguchi}, {Lamareille}, {Leauthaud}, {Sasaki}, {Thompson},
  {Zamojski}, {Zamorani}, {Bardelli}, {Bolzonella}, {Bongiorno}, {Brusa},
  {Caputi}, {Carollo}, {Contini}, {Cook}, {Coppa}, {Cucciati}, {de la Torre},
  {de Ravel}, {Franzetti}, {Garilli}, {Hasinger}, {Iovino}, {Kampczyk},
  {Kneib}, {Knobel}, {Kovac}, {LeBorgne}, {LeBrun}, {LeF{\`e}vre}, {Lilly},
  {Looper}, {Maier}, {Mainieri}, {Mellier}, {Mignoli}, {Murayama}, {Pell{\`o}},
  {Peng}, {P{\'e}rez-Montero}, {Renzini}, {Ricciardelli}, {Schiminovich},
  {Scodeggio}, {Shioya}, {Silverman}, {Surace}, {Tanaka}, {Tasca}, {Tresse},
  {Vergani}, \& {Zucca}}]{Ilbert2009}
{Ilbert}, O., {Capak}, P., {Salvato}, M., {et~al.} 2009, \apj, 690, 1236

\bibitem[{{Kauffmann} {et~al.}(2004){Kauffmann}, {White}, {Heckman},
  {M{\'e}nard}, {Brinchmann}, {Charlot}, {Tremonti}, \&
  {Brinkmann}}]{Kauffmann2004}
{Kauffmann}, G., {White}, S.~D.~M., {Heckman}, T.~M., {et~al.} 2004, \mnras,
  353, 713

\bibitem[{{Kawata} \& {Mulchaey}(2008)}]{Kawata2008}
{Kawata}, D. \& {Mulchaey}, J.~S. 2008, \apjl, 672, L103

\bibitem[{{Kenney} {et~al.}(2004){Kenney}, {van Gorkom}, \&
  {Vollmer}}]{Kenney2004}
{Kenney}, J.~D.~P., {van Gorkom}, J.~H., \& {Vollmer}, B. 2004, \aj, 127, 3361

\bibitem[{{Kinney} {et~al.}(1996){Kinney}, {Calzetti}, {Bohlin}, {McQuade},
  {Storchi-Bergmann}, \& {Schmitt}}]{Kinney1996}
{Kinney}, A.~L., {Calzetti}, D., {Bohlin}, R.~C., {et~al.} 1996, \apj, 467, 38

\bibitem[{{Kitzbichler} \& {White}(2007)}]{KitzbichlerWhite2007}
{Kitzbichler}, M.~G. \& {White}, S.~D.~M. 2007, \mnras, 376, 2

\bibitem[{{Knobel} {et~al.}(2009){Knobel}, {Lilly}, {Iovino}, {Porciani},
  {Kova{\v c}}, {Cucciati}, {Finoguenov}, {Kitzbichler}, {Carollo}, {Contini},
  {Kneib}, {LeF{\`e}vre}, {Mainieri}, {Renzini}, {Scodeggio}, {Zamorani},
  {Bardelli}, {Bolzonella}, {Bongiorno}, {Caputi}, {Coppa}, {de la Torre}, {de
  Ravel}, {Franzetti}, {Garilli}, {Kampczyk}, {Lamareille}, {LeBorgne},
  {LeBrun}, {Maier}, {Mignoli}, {Pello}, {Peng}, {Montero}, {Ricciardelli},
  {Silverman}, {Tanaka}, {Tasca}, {Tresse}, {Vergani}, {Zucca}, {Abbas},
  {Bottini}, {Cappi}, {Cassata}, {Cimatti}, {Fumana}, {Guzzo}, {Koekemoer},
  {Leauthaud}, {Maccagni}, {Marinoni}, {McCracken}, {Memeo}, {Meneux}, {Oesch},
  {Pozzetti}, \& {Scaramella}}]{Knobel2009}
{Knobel}, C., {Lilly}, S.~J., {Iovino}, A., {et~al.} 2009, \apj, 697, 1842

\bibitem[{{Kodama} \& {Bower}(2001)}]{Kodama2001}
{Kodama}, T. \& {Bower}, R.~G. 2001, \mnras, 321, 18

\bibitem[{{Koekemoer} {et~al.}(2007){Koekemoer}, {Aussel}, {Calzetti}, {Capak},
  {Giavalisco}, {Kneib}, {Leauthaud}, {Le F{\`e}vre}, {McCracken}, {Massey},
  {Mobasher}, {Rhodes}, {Scoville}, \& {Shopbell}}]{Koekemoer2007}
{Koekemoer}, A.~M., {Aussel}, H., {Calzetti}, D., {et~al.} 2007, \apjs, 172,
  196

\bibitem[{{Kova{\v c}} {et~al.}(2009{\natexlab{a}}){Kova{\v c}}, {Lilly},
  {Cucciati}, {Porciani}, {Iovino}, {Zamorani}, {Oesch}, {Bolzonella},
  {Knobel}, {Finoguenov}, {Peng}, {Carollo}, {Pozzetti}, {Caputi}, {Silverman},
  {Tasca}, {Scodeggio}, {Vergani}, {Zucca}, {Contini}, {Kneib}, {Le Fevre},
  {Mainieri}, {Renzini}, {Scoville}, {Capak}, {Bardelli}, {Bongiorno}, {Coppa},
  {de la Torre}, {de Ravel}, {Franzetti}, {Garilli}, {Guzzo}, {Kampczyk},
  {Lamareille}, {Le Borgne}, {Le Brun}, {Maier}, {Mignoli}, {Pello}, {Perez
  Montero}, {Ricciardelli}, {Tanaka}, {Tresse}, {Abbas}, {Bottini}, {Cappi},
  {Cassata}, {Cimatti}, {Fumana}, {Maccagni}, {Marinoni}, {McCracken}, {Memeo},
  {Meneux}, {Scaramella}, \& {Koekemoer}}]{Kovac2009a}
{Kova{\v c}}, K., {Lilly}, S.~J., {Cucciati}, O., {et~al.} 2009{\natexlab{a}},
  ArXiv e-prints 0903.3409

\bibitem[{{Kova{\v c}} {et~al.}(2009{\natexlab{b}}){Kova{\v c}}, {Lilly},
  {Knobel}, {Pinco}, {Pallo}, {Pinco}, \& {Pallo}}]{Kovac2009b}
{Kova{\v c}}, K., {Lilly}, S.~J., {Knobel}, C., {et~al.} 2009{\natexlab{b}},
  \apj, submitted

\bibitem[{{Koyama} {et~al.}(2007){Koyama}, {Kodama}, {Tanaka}, {Shimasaku}, \&
  {Okamura}}]{Koyama2007}
{Koyama}, Y., {Kodama}, T., {Tanaka}, M., {Shimasaku}, K., \& {Okamura}, S.
  2007, \mnras, 382, 1719

\bibitem[{{Kroupa}(2001)}]{Kroupa2001}
{Kroupa}, P. 2001, \mnras, 322, 231

\bibitem[{{Larson} {et~al.}(1980){Larson}, {Tinsley}, \&
  {Caldwell}}]{Larson1980}
{Larson}, R.~B., {Tinsley}, B.~M., \& {Caldwell}, C.~N. 1980, \apj, 237, 692

\bibitem[{{Lavery} \& {Henry}(1986)}]{Lavery1986}
{Lavery}, R.~J. \& {Henry}, J.~P. 1986, \apjl, 304, L5

\bibitem[{{Lavery} \& {Henry}(1988)}]{Lavery1988}
{Lavery}, R.~J. \& {Henry}, J.~P. 1988, \apj, 330, 596

\bibitem[{{Le Fevre} {et~al.}(1995){Le Fevre}, {Crampton}, {Lilly}, {Hammer},
  \& {Tresse}}]{LeFevre1995}
{Le Fevre}, O., {Crampton}, D., {Lilly}, S.~J., {Hammer}, F., \& {Tresse}, L.
  1995, \apj, 455, 60

\bibitem[{{Le F{\`e}vre} {et~al.}(2005){Le F{\`e}vre}, {Vettolani}, {Garilli},
  {Tresse}, {Bottini}, {Le Brun}, {Maccagni}, {Picat}, {Scaramella},
  {Scodeggio}, {Zanichelli}, {Adami}, {Arnaboldi}, {Arnouts}, {Bardelli},
  {Bolzonella}, {Cappi}, {Charlot}, {Ciliegi}, {Contini}, {Foucaud},
  {Franzetti}, {Gavignaud}, {Guzzo}, {Ilbert}, {Iovino}, {McCracken}, {Marano},
  {Marinoni}, {Mathez}, {Mazure}, {Meneux}, {Merighi}, {Paltani}, {Pell{\`o}},
  {Pollo}, {Pozzetti}, {Radovich}, {Zamorani}, {Zucca}, {Bondi}, {Bongiorno},
  {Busarello}, {Lamareille}, {Mellier}, {Merluzzi}, {Ripepi}, \&
  {Rizzo}}]{LeFevre2005}
{Le F{\`e}vre}, O., {Vettolani}, G., {Garilli}, B., {et~al.} 2005, \aap, 439,
  845

\bibitem[{{Lilly} {et~al.}(1996){Lilly}, {Le Fevre}, {Hammer}, \&
  {Crampton}}]{Lilly1996}
{Lilly}, S.~J., {Le Fevre}, O., {Hammer}, F., \& {Crampton}, D. 1996, \apjl,
  460, L1+

\bibitem[{{Lilly} {et~al.}(2007){Lilly}, {Le F{\`e}vre}, {Renzini}, {Zamorani},
  {Scodeggio}, {Contini}, {Carollo}, {Hasinger}, {Kneib}, {Iovino}, {Le Brun},
  {Maier}, {Mainieri}, {Mignoli}, {Silverman}, {Tasca}, {Bolzonella},
  {Bongiorno}, {Bottini}, {Capak}, {Caputi}, {Cimatti}, {Cucciati}, {Daddi},
  {Feldmann}, {Franzetti}, {Garilli}, {Guzzo}, {Ilbert}, {Kampczyk}, {Kovac},
  {Lamareille}, {Leauthaud}, {Borgne}, {McCracken}, {Marinoni}, {Pello},
  {Ricciardelli}, {Scarlata}, {Vergani}, {Sanders}, {Schinnerer}, {Scoville},
  {Taniguchi}, {Arnouts}, {Aussel}, {Bardelli}, {Brusa}, {Cappi}, {Ciliegi},
  {Finoguenov}, {Foucaud}, {Franceschini}, {Halliday}, {Impey}, {Knobel},
  {Koekemoer}, {Kurk}, {Maccagni}, {Maddox}, {Marano}, {Marconi}, {Meneux},
  {Mobasher}, {Moreau}, {Peacock}, {Porciani}, {Pozzetti}, {Scaramella},
  {Schiminovich}, {Shopbell}, {Smail}, {Thompson}, {Tresse}, {Vettolani},
  {Zanichelli}, \& {Zucca}}]{Lilly2007}
{Lilly}, S.~J., {Le F{\`e}vre}, O., {Renzini}, A., {et~al.} 2007, \apjs, 172,
  70

\bibitem[{{Lilly et al.}(2009)}]{Lilly2009}
{Lilly et al.} 2009, submitted

\bibitem[{{Madau} {et~al.}(1998){Madau}, {Pozzetti}, \&
  {Dickinson}}]{Madau1998}
{Madau}, P., {Pozzetti}, L., \& {Dickinson}, M. 1998, \apj, 498, 106

\bibitem[{{Margoniner} \& {de Carvalho}(2000)}]{Margoniner2000}
{Margoniner}, V.~E. \& {de Carvalho}, R.~R. 2000, \aj, 119, 1562

\bibitem[{{Margoniner} {et~al.}(2001){Margoniner}, {de Carvalho}, {Gal}, \&
  {Djorgovski}}]{Margoniner2001}
{Margoniner}, V.~E., {de Carvalho}, R.~R., {Gal}, R.~R., \& {Djorgovski}, S.~G.
  2001, \apjl, 548, L143

\bibitem[{{Marinoni} {et~al.}(2002){Marinoni}, {Davis}, {Newman}, \&
  {Coil}}]{Marinoni2002}
{Marinoni}, C., {Davis}, M., {Newman}, J.~A., \& {Coil}, A.~L. 2002, \apj, 580,
  122

\bibitem[{{Mart{\'{\i}}nez} {et~al.}(2002){Mart{\'{\i}}nez}, {Zandivarez},
  {Dom{\'{\i}}nguez}, {Merch{\'a}n}, \& {Lambas}}]{Martinez2002a}
{Mart{\'{\i}}nez}, H.~J., {Zandivarez}, A., {Dom{\'{\i}}nguez}, M.,
  {Merch{\'a}n}, M.~E., \& {Lambas}, D.~G. 2002, \mnras, 333, L31

\bibitem[{{Mignoli} {et~al.}(2009){Mignoli}, {Zamorani}, {Scodeggio},
  {Cimatti}, {Halliday}, {Lilly}, {Pozzetti}, {Vergani}, {Carollo}, {Contini},
  {Le F{\'e}vre}, {Mainieri}, {Renzini}, {Bardelli}, {Bolzonella}, {Bongiorno},
  {Caputi}, {Coppa}, {Cucciati}, {de La Torre}, {de Ravel}, {Franzetti},
  {Garilli}, {Iovino}, {Kampczyk}, {Kneib}, {Knobel}, {Kova{\v c}},
  {Lamareille}, {Le Borgne}, {Le Brun}, {Maier}, {Pell{\`o}}, {Peng}, {Perez
  Montero}, {Ricciardelli}, {Scarlata}, {Silverman}, {Tanaka}, {Tasca},
  {Tresse}, {Zucca}, {Abbas}, {Bottini}, {Capak}, {Cappi}, {Cassata}, {Fumana},
  {Guzzo}, {Leauthaud}, {Maccagni}, {Marinoni}, {McCracken}, {Memeo}, {Meneux},
  {Oesch}, {Porciani}, {Scaramella}, \& {Scoville}}]{Mignoli2009}
{Mignoli}, M., {Zamorani}, G., {Scodeggio}, M., {et~al.} 2009, \aap, 493, 39

\bibitem[{{Moore} {et~al.}(1999){Moore}, {Lake}, {Quinn}, \&
  {Stadel}}]{Moore1999}
{Moore}, B., {Lake}, G., {Quinn}, T., \& {Stadel}, J. 1999, \mnras, 304, 465

\bibitem[{{Poggianti} {et~al.}(1999){Poggianti}, {Smail}, {Dressler}, {Couch},
  {Barger}, {Butcher}, {Ellis}, \& {Oemler}}]{Poggianti1999}
{Poggianti}, B.~M., {Smail}, I., {Dressler}, A., {et~al.} 1999, \apj, 518, 576

\bibitem[{{Poggianti} {et~al.}(2006){Poggianti}, {von der Linden}, {De Lucia},
  {Desai}, {Simard}, {Halliday}, {Arag{\'o}n-Salamanca}, {Bower}, {Varela},
  {Best}, {Clowe}, {Dalcanton}, {Jablonka}, {Milvang-Jensen}, {Pello},
  {Rudnick}, {Saglia}, {White}, \& {Zaritsky}}]{Poggianti2006}
{Poggianti}, B.~M., {von der Linden}, A., {De Lucia}, G., {et~al.} 2006, \apj,
  642, 188

\bibitem[{{Popesso} {et~al.}(2007){Popesso}, {Biviano}, {Romaniello}, \&
  {B{\"o}hringer}}]{Popesso2007}
{Popesso}, P., {Biviano}, A., {Romaniello}, M., \& {B{\"o}hringer}, H. 2007,
  \aap, 461, 411

\bibitem[{{Pozzetti} {et~al.}(2007){Pozzetti}, {Bolzonella}, {Lamareille},
  {Zamorani}, {Franzetti}, {Le F{\`e}vre}, {Iovino}, {Temporin}, {Ilbert},
  {Arnouts}, {Charlot}, {Brinchmann}, {Zucca}, {Tresse}, {Scodeggio}, {Guzzo},
  {Bottini}, {Garilli}, {Le Brun}, {Maccagni}, {Picat}, {Scaramella},
  {Vettolani}, {Zanichelli}, {Adami}, {Bardelli}, {Cappi}, {Ciliegi},
  {Contini}, {Foucaud}, {Gavignaud}, {McCracken}, {Marano}, {Marinoni},
  {Mazure}, {Meneux}, {Merighi}, {Paltani}, {Pell{\`o}}, {Pollo}, {Radovich},
  {Bondi}, {Bongiorno}, {Cucciati}, {de la Torre}, {Gregorini}, {Mellier},
  {Merluzzi}, {Vergani}, \& {Walcher}}]{Pozzetti2007}
{Pozzetti}, L., {Bolzonella}, M., {Lamareille}, F., {et~al.} 2007, \aap, 474,
  443

\bibitem[{{Pozzetti} {et~al.}(2009){Pozzetti}, {Bolzonella}, {Zucca},
  {Zamorani}, {Lilly}, {Renzini}, {Moresco}, {Mignoli}, {Cassata}, {Tasca},
  {Lamareille}, {Maier}, {Meneux}, {Oesch}, {Vergani}, {Caputi}, {Kova{\v c}},
  {Cimatti}, {Cucciati}, {Iovino}, {Peng}, {Carollo}, {Contini}, {Kneib}, {Le
  Fevre}, {Mainieri}, {Scodeggio}, {Bardelli}, {Bongiorno}, {Coppa}, {.~de la
  Torre}, {de Ravel}, {Franzetti}, {Garilli}, {Kampczyk}, {Knobel}, {Le
  Borgne}, {Le Brun}, {Pello}, {Perez Montero}, {Ricciardelli}, {Silverman},
  {Tanaka}, {Tresse}, {Abbas}, {Bottini}, {Cappi}, {Guzzo}, {Halliday},
  {Leauthaud}, {Koekemoer}, {Maccagni}, {Marinoni}, {McCracken}, {Memeo},
  {Porciani}, {Scaramella}, {Scarlata}, \& {Scoville}}]{Pozzetti2009}
{Pozzetti}, L., {Bolzonella}, M., {Zucca}, E., {et~al.} 2009, ArXiv e-prints
  0907.5416

\bibitem[{{Rakos} \& {Schombert}(1995)}]{Rakos1995}
{Rakos}, K.~D. \& {Schombert}, J.~M. 1995, \apj, 439, 47

\bibitem[{{Roediger} \& {Hensler}(2005)}]{Roediger2005}
{Roediger}, E. \& {Hensler}, G. 2005, \aap, 433, 875

\bibitem[{{Schiminovich} {et~al.}(2005){Schiminovich}, {Ilbert}, {Arnouts},
  {Milliard}, {Tresse}, {Le F{\`e}vre}, {Treyer}, {Wyder}, {Budav{\'a}ri},
  {Zucca}, {Zamorani}, {Martin}, {Adami}, {Arnaboldi}, {Bardelli}, {Barlow},
  {Bianchi}, {Bolzonella}, {Bottini}, {Byun}, {Cappi}, {Contini}, {Charlot},
  {Donas}, {Forster}, {Foucaud}, {Franzetti}, {Friedman}, {Garilli},
  {Gavignaud}, {Guzzo}, {Heckman}, {Hoopes}, {Iovino}, {Jelinsky}, {Le Brun},
  {Lee}, {Maccagni}, {Madore}, {Malina}, {Marano}, {Marinoni}, {McCracken},
  {Mazure}, {Meneux}, {Morrissey}, {Neff}, {Paltani}, {Pell{\`o}}, {Picat},
  {Pollo}, {Pozzetti}, {Radovich}, {Rich}, {Scaramella}, {Scodeggio},
  {Seibert}, {Siegmund}, {Small}, {Szalay}, {Vettolani}, {Welsh}, {Xu}, \&
  {Zanichelli}}]{Schiminovich2005}
{Schiminovich}, D., {Ilbert}, O., {Arnouts}, S., {et~al.} 2005, \apjl, 619, L47

\bibitem[{{Scoville} {et~al.}(2007){Scoville}, {Aussel}, {Brusa}, {Capak},
  {Carollo}, {Elvis}, {Giavalisco}, {Guzzo}, {Hasinger}, {Impey}, {Kneib},
  {LeFevre}, {Lilly}, {Mobasher}, {Renzini}, {Rich}, {Sanders}, {Schinnerer},
  {Schminovich}, {Shopbell}, {Taniguchi}, \& {Tyson}}]{Scoville2007}
{Scoville}, N., {Aussel}, H., {Brusa}, M., {et~al.} 2007, \apjs, 172, 1

\bibitem[{{Silverman} {et~al.}(2009){Silverman}, {Kova{\v c}}, {Knobel},
  {Lilly}, {Bolzonella}, {Lamareille}, {Mainieri}, {Brusa}, {Cappelluti},
  {Peng}, {Hasinger}, {Zamorani}, {Scodeggio}, {Contini}, {Carollo}, {Jahnke},
  {Kneib}, {LeFevre}, {Bardelli}, {Bongiorno}, {Brunner}, {Caputi}, {Civano},
  {Comastri}, {Coppa}, {Cucciati}, {de la Torre}, {de Ravel}, {Elvis},
  {Finoguenov}, {Fiore}, {Franzetti}, {Garilli}, {Gilli}, {Griffiths},
  {Iovino}, {Kampczyk}, {Koekemoer}, {LeBorgne}, {LeBrun}, {Maier}, {Mignoli},
  {Pello}, {Perez Montero}, {Ricciardelli}, {Tanaka}, {Tasca}, {Tresse},
  {Vergani}, {Vignali}, {Zucca}, {Bottini}, {Cappi}, {Cassata}, {Marinoni},
  {McCracken}, {Memeo}, {Meneux}, {Oesch}, {Porciani}, \&
  {Salvato}}]{Silverman2009}
{Silverman}, J.~D., {Kova{\v c}}, K., {Knobel}, C., {et~al.} 2009, \apj, 695,
  171

\bibitem[{{Smail} {et~al.}(1998){Smail}, {Edge}, {Ellis}, \&
  {Blandford}}]{Smail1998}
{Smail}, I., {Edge}, A.~C., {Ellis}, R.~S., \& {Blandford}, R.~D. 1998, \mnras,
  293, 124

\bibitem[{{Springel}(2005)}]{Springel2005}
{Springel}, V. 2005, \mnras, 364, 1105

\bibitem[{{Strateva} {et~al.}(2001){Strateva}, {Ivezi{\'c}}, {Knapp},
  {Narayanan}, {Strauss}, {Gunn}, {Lupton}, {Schlegel}, {Bahcall}, {Brinkmann},
  {Brunner}, {Budav{\'a}ri}, {Csabai}, {Castander}, {Doi}, {Fukugita}, {Gy{\H
  o}ry}, {Hamabe}, {Hennessy}, {Ichikawa}, {Kunszt}, {Lamb}, {McKay},
  {Okamura}, {Racusin}, {Sekiguchi}, {Schneider}, {Shimasaku}, \&
  {York}}]{Strateva2001}
{Strateva}, I., {Ivezi{\'c}}, {\v Z}., {Knapp}, G.~R., {et~al.} 2001, \aj, 122,
  1861

\bibitem[{{Tanaka} {et~al.}(2008){Tanaka}, {Finoguenov}, {Kodama}, {Morokuma},
  {Rosati}, {Stanford}, {Eisenhardt}, {Holden}, \& {Mei}}]{Tanaka2008}
{Tanaka}, M., {Finoguenov}, A., {Kodama}, T., {et~al.} 2008, \aap, 489, 571

\bibitem[{{Tanaka} {et~al.}(2004){Tanaka}, {Goto}, {Okamura}, {Shimasaku}, \&
  {Brinkmann}}]{Tanaka2004}
{Tanaka}, M., {Goto}, T., {Okamura}, S., {Shimasaku}, K., \& {Brinkmann}, J.
  2004, \aj, 128, 2677

\bibitem[{{Tasca} {et~al.}(2009){Tasca}, {Kneib}, {Iovino}, {Le F{\`e}vre},
  {Kova{\v c}}, {Bolzonella}, {Lilly}, {Abraham}, {Cassata}, {Cucciati},
  {Guzzo}, {Tresse}, {Zamorani}, {Capak}, {Garilli}, {Scodeggio}, {Sheth},
  {Zucca}, {Carollo}, {Contini}, {Mainieri}, {Renzini}, {Bardelli},
  {Bongiorno}, {Caputi}, {Coppa}, {de La Torre}, {de Ravel}, {Franzetti},
  {Kampczyk}, {Knobel}, {Koekemoer}, {Lamareille}, {Le Borgne}, {Le Brun},
  {Maier}, {Mignoli}, {Pello}, {Peng}, {Perez Montero}, {Ricciardelli},
  {Silverman}, {Vergani}, {Tanaka}, {Abbas}, {Bottini}, {Cappi}, {Cimatti},
  {Ilbert}, {Leauthaud}, {Maccagni}, {Marinoni}, {McCracken}, {Memeo},
  {Meneux}, {Oesch}, {Porciani}, {Pozzetti}, {Scaramella}, \&
  {Scarlata}}]{Tasca2009}
{Tasca}, L.~A.~M., {Kneib}, J.-P., {Iovino}, A., {et~al.} 2009, \aap, 503, 379

\bibitem[{{van Dokkum} {et~al.}(2000){van Dokkum}, {Franx}, {Fabricant},
  {Illingworth}, \& {Kelson}}]{vanDokkum2000}
{van Dokkum}, P.~G., {Franx}, M., {Fabricant}, D., {Illingworth}, G.~D., \&
  {Kelson}, D.~D. 2000, \apj, 541, 95

\bibitem[{{Vergani} {et~al.}(2009){Vergani}, {Zamorani}, {Lilly}, {Pinco},
  {Pallo}, {Pinco}, \& {Pallo}}]{Vergani2009}
{Vergani}, D., {Zamorani}, G., {Lilly}, S.~J., {et~al.} 2009, \aap, submitted

\bibitem[{{Vollmer} {et~al.}(2004){Vollmer}, {Beck}, {Kenney}, \& {van
  Gorkom}}]{Vollmer2004}
{Vollmer}, B., {Beck}, R., {Kenney}, J.~D.~P., \& {van Gorkom}, J.~H. 2004,
  \aj, 127, 3375

\bibitem[{{Voronoi}(1908)}]{Voronoi1908}
{Voronoi}, G. 1908, J. Reine Angew. Math., 134, 198

\bibitem[{{Wechsler} {et~al.}(2002){Wechsler}, {Bullock}, {Primack},
  {Kravtsov}, \& {Dekel}}]{Wechsler2002}
{Wechsler}, R.~H., {Bullock}, J.~S., {Primack}, J.~R., {Kravtsov}, A.~V., \&
  {Dekel}, A. 2002, \apj, 568, 52

\bibitem[{{Weinmann} {et~al.}(2000){Weinmann}, {PP}, {GG}, {Illingworth}, \&
  {Kelson}}]{Weinmann06}
{Weinmann}, P.~G., {PP}, M., {GG}, D., {Illingworth}, G.~D., \& {Kelson}, D.~D.
  2000, \apj, 541, 95

\bibitem[{{Wilman} {et~al.}(2005{\natexlab{a}}){Wilman}, {Balogh}, {Bower},
  {Mulchaey}, {Oemler}, {Carlberg}, {Eke}, {Lewis}, {Morris}, \&
  {Whitaker}}]{Wilman2005b}
{Wilman}, D.~J., {Balogh}, M.~L., {Bower}, R.~G., {et~al.} 2005{\natexlab{a}},
  \mnras, 358, 88

\bibitem[{{Wilman} {et~al.}(2005{\natexlab{b}}){Wilman}, {Balogh}, {Bower},
  {Mulchaey}, {Oemler}, {Carlberg}, {Morris}, \& {Whitaker}}]{Wilman2005a}
{Wilman}, D.~J., {Balogh}, M.~L., {Bower}, R.~G., {et~al.} 2005{\natexlab{b}},
  \mnras, 358, 71

\bibitem[{{Zabludoff} \& {Mulchaey}(1998)}]{ZabludoffMulchaey1998a}
{Zabludoff}, A.~I. \& {Mulchaey}, J.~S. 1998, \apj, 496, 39

\bibitem[{{Zucca} {et~al.}(2009){Zucca}, {Bardelli}, {Bolzonella}, {Pinco},
  {Pallo}, {Pinco}, \& {Pallo}}]{Zucca2009}
{Zucca}, E., {Bardelli}, S., {Bolzonella}, M., {et~al.} 2009, \aap, submitted

\end{thebibliography}

\end{document}